%% file: scifile_Main.tex
\documentclass[12pt]{article}

\usepackage{scicite}
\usepackage{times}
\usepackage{hyperref}
\usepackage{amssymb}
\usepackage{enumitem}
\usepackage{mathtools}
\usepackage{graphicx}
\usepackage{amssymb}
\usepackage{amsmath}
\usepackage{ulem}
\usepackage{epstopdf}
\usepackage{color}
\usepackage{graphicx}
\usepackage{xcolor}
\usepackage{textcomp}

\def\aap{Astron. Astrophys.}                
\newcommand{\aj}{Astron. J.}
\newcommand{\apjs}{Astrophys. J.}
\newcommand{\apj}{Astrophys. J.}
\newcommand{\apjl}{Astrophys. J.}
\newcommand{\mnras}{Mon. Not. R. Astron. Soc.}

\newcommand{\nat}{Nature}

\newcommand{\procspie}{Proc. SPIE.}
\newcommand{\pasp}{Publ. Astron. Soc. Pac.}

\newcommand{\pasj}{Publ. Astron. Soc. Jpn.}
\newcommand{\ropp}{Rep. Prog. Phys.}

\newcommand{\red}[1]{\textcolor{black}{#1}}

\newcommand{\green}[1]{\textcolor{black}{}} 
\newcommand{\MYsout}[1]{\sout{}}

\hypersetup{
    colorlinks=true,
    linkcolor=blue,
    filecolor=magenta,      
    urlcolor=cyan,
}
\usepackage{url}
\usepackage{listings}
\lstdefinestyle{mystyle}{
  commentstyle=\color{codegreen},
  keywordstyle=\color{magenta},
  numberstyle=\tiny\color{codegray},
  stringstyle=\color{codepurple},
  basicstyle=\footnotesize,
  breakatwhitespace=false,         
  breaklines=true,                 
  captionpos=b,                    
  keepspaces=true,                 
  numbers=left,                    
  numbersep=5pt,                  
  showspaces=false,                
  showstringspaces=false,
  showtabs=false,                  
  tabsize=2
}
\usepackage{lineno}

\usepackage{graphicx}

\usepackage[affil-it]{authblk} 

\newenvironment{sciabstract}{%
\begin{quote} \bf}
{\end{quote}}

\newcounter{lastnote}

\title{\red{Enhanced} X-ray \red{Emission} Coinciding with Giant Radio Pulses from the Crab Pulsar}

\author[1$\ast\dagger$]{Teruaki Enoto}
\author[2,3,4$\ast\dagger$]{Toshio Terasawa} 
\author[5,6,7$\ast\dagger$]{Shota Kisaka}
\author[1,8,9$\ast\dagger$]{Chin-Ping Hu}
\author[10]{Sebastien Guillot}
\author[11]{Natalia Lewandowska}
\author[12,13]{Christian Malacaria}
\author[14]{Paul S. Ray}
\author[11,15]{Wynn C.G. Ho}
\author[16,17]{Alice K. Harding}
\author[16]{Takashi Okajima}
\author[16]{Zaven Arzoumanian}
\author[16]{Keith C. Gendreau}
\author[16,18]{Zorawar Wadiasingh}
\author[16]{Craig B. Markwardt}
\author[16]{Yang Soong}
\author[16]{Steve Kenyon}
\author[19]{Slavko Bogdanov}
\author[20,21]{Walid A. Majid}
\author[22]{Tolga G\"uver}
\author[23]{Gaurava K. Jaisawal}
\author[24]{Rick Foster}
\author[25,26,27]{Yasuhiro Murata}
\author[25,27]{Hiroshi Takeuchi}
\author[26,28]{Kazuhiro Takefuji}
\author[28]{Mamoru Sekido}
\author[29]{Yoshinori Yonekura}
\author[30]{Hiroaki Misawa}
\author[30]{Fuminori Tsuchiya}
\author[31]{Takahiko Aoki}
\author[32]{Munetoshi Tokumaru}
\author[33,34,35]{Mareki Honma}
\author[33,35]{Osamu Kameya}
\author[33]{Tomoaki Oyama}
\author[2]{Katsuaki Asano}
\author[36]{Shinpei Shibata}
\author[37]{Shuta J. Tanaka}

\affil[1]{Cluster for Pioneering Research, RIKEN, Wako 351-0198, Japan}
\affil[2]{Institute for Cosmic Ray Research, University of Tokyo, Kashiwa 277-8582, Japan}
\affil[3]{Mizusawa VLBI Observatory, National Astronomical Observatory of Japan, Mitaka 181-8588, Japan}
\affil[4]{Interdisciplinary Theoretical Science Research Group, RIKEN, Wako 351-0198, Japan}
\affil[5]{Frontier Research Institute for Interdisciplinary Sciences, Tohoku University, Sendai 980-8578, Japan}
\affil[6]{Astronomical Institute, Tohoku University, Sendai 980-8578, Japan}
\affil[7]{Department of Physical Science, Hiroshima University, Higashi-Hiroshima 739-8526, Japan}
\affil[8]{Department of Physics, National Changhua University of Education, Changhua 50007, Taiwan}
\affil[9]{Department of Astronomy, Kyoto University, Kyoto 606-8502, Japan}
\affil[10]{Institut de Recherche en Astrophysique, Toulouse, 31028, France}
\affil[11]{Department of Physics and Astronomy, Haverford College, Haverford, PA, 19041, USA}
\affil[12]{NASA Marshall Space Flight Center, National Space Science and Technology Center, Huntsville, AL 35805, USA}
\affil[13]{Universities Space Research Association, Science and Technology Institute, Huntsville, AL 35805, USA}
\affil[14]{U.S. Naval Research Laboratory, Washington DC 20375, USA}
\affil[15]{Mathematical Sciences and STAG Research Centre, University of Southampton, Southampton, SO17 1BJ, United Kingdom}
\affil[16]{NASA Goddard Space Flight Center, Greenbelt, MD 20771, USA}
\affil[17]{Theoretical Division, Los Alamos National Laboratory, Los Alamos, NM 87545, USA}
\affil[18]{Universities Space Research Association, Columbia, MD 21046, USA}
\affil[19]{Columbia Astrophysics Laboratory, Columbia University, New York, NY 10027, USA}
\affil[20]{Jet Propulsion Laboratory, California Institute of Technology, Pasadena, CA 91109, USA}
\affil[21]{California Institute of Technology, Pasadena, CA 91125, USA}
\affil[22]{Istanbul University, Science Faculty, Department of Astronomy and Space Sciences, Beyaz\i t, 34119, Istanbul, Turkey}
\affil[23]{National Space Institute, Technical University of Denmark, Elektrovej 327-328, Denmark}
\affil[24]{Massachusetts Institute of Technology Kavli Institute for Astrophysics and Space Research, 77 Massachusetts Avenue, Cambridge, MA 02139}
\affil[25]{Institute of Space and Astronautical Science, Japan Aerospace Exploration Agency, Sagamihara 252-5210, Japan}
\affil[26]{Usuda Deep Space Center, Japan Aerospace Exploration Agency, Saku 384-0306, Japan}
\affil[27]{Department of Space and Astronautical Science, SOKENDAI (The Graduate University for Advanced Studies), Sagamihara 252-5210, Japan}
\affil[28]{Kashima Space Technology Center, National Institute of Information and Communications Technology, Kashima 314-8501, Japan}
\affil[29]{Center for Astronomy, Ibaraki University, Mito 310-8512, Japan}
\affil[30]{Planetary Plasma and Atmospheric Research Center, Tohoku University, Sendai 980-8578, Japan}
\affil[31]{The Research Institute for Time Studies, Yamaguchi University, Yamaguchi 753-8511, Japan}
\affil[32]{Institute for Space-Earth Environmental Research, Nagoya University, Nagoya 464-8601, Japan}
\affil[33]{Mizusawa VLBI Observatory, National Astronomical Observatory of Japan, Oshu 023-0861, Japan}
\affil[34]{Department of Astronomy, University of Tokyo, Tokyo 113-0033, Japan}
\affil[35]{Department of Astronomical Science, SOKENDAI (The Graduate University for Advanced Studies), Mitaka 181-8588, Japan}
\affil[36]{Department of Physics, Yamagata University, Yamagata 990-8560, Japan}
\affil[37]{Department of Physics and Mathematics, Aoyama Gakuin University, Sagamihara, 252-5258, Japan}

\affil[$\ast$]{These authors contributed equally to this work.}
\affil[$\dagger$]{Corresponding author. E-mail: teruaki.enoto@riken.jp (T.E.); terasawa@icrr.u-tokyo.ac.jp (T.T.); kisaka@hiroshima-u.ac.jp (S.K.); cphu0821@cc.ncue.edu.tw (C.-P. H.)}

\date{}

\begin{document} 

\baselineskip24pt

\maketitle 

\begin{sciabstract}
Giant radio pulses (GRPs) are sporadic bursts \red{emitted by some pulsars, lasting a few microseconds. GRPs are} hundreds to thousands of times brighter than regular pulses \red{from these sources. The only GRP-associated emission outside radio wavelengths is} from the Crab \red{Pulsar}, where \red{optical emission is enhanced by a few percent during GRPs. We observed the Crab Pulsar simultaneously at X-ray and radio wavelengths, finding enhancement of the X-ray emission by $3.8\pm0.7\%$ (a 5.4$\sigma$ detection) coinciding with GRPs.
This implies} that the total emitted energy from \red{GRPs} is tens to hundreds of times higher than previously known. \red{We discuss the} implications for the pulsar emission mechanism and extragalactic fast radio bursts.
\end{sciabstract}

\vspace{1cm}

Spinning neutron stars \red{emit periodic radio pulses from their magnetospheres, which can be observed as a pulsar.} The radio pulses \red{are emitted by} a coherent mechanism \red{\cite{2016JPlPh..82b6302M}}. Some pulsars also show optical, X-ray, and gamma-ray pulses, \red{which are usually interpreted using incoherent emission mechanisms.} Giant Radio Pulses (GRPs) are a form of sporadic pulsar emission with radio \red{fluences} at least an order of magnitude \red{higher} than those of regular pulses, \red{of unknown origin} \cite{2006ApJ...640..941K,2012ApJ...749...24B}. \red{GRPs have} a power-law intensity distribution, \red{unlike regular pulses which have log-normal} or exponential intensity distributions \cite{2011ApJ...741...53M}. 

GRPs are bright, sometimes exceeding \red{a megajansky (MJy, $1~{\rm Jy} = 10^{-26}~{\rm W~m}^{-2}~{\rm Hz}^{-1}$) for a few} nano- to microseconds \cite{2007ApJ...670..693H}. \red{This is sufficient to detect each GRP} during a single stellar rotation. GRPs from young neutron stars \red{have been} proposed \red{as} the origin of \red{fast radio bursts} (FRBs), short-duration radio transients at cosmological distances \cite{2007Sci...318..777L,2016MNRAS.457..232C}. \red{A nearby counterpart has provided evidence for this association} \cite{2020Natur.587...59B, 2020Natur.587...54C} and for a connection between coherent and incoherent processes in the neutron star magnetosphere \red{\cite{2020MNRAS.498.1397L}}. Although GRPs are not the leading explanation for FRBs, the broad-band characteristics of GRPs \red{provide information on} coherent radio emission in neutron star \red{magnetospheres}, which \red{may be relevant to} FRBs.

GRPs have been detected from only a small fraction of pulsars \cite{2005AJ....129.1993M,Enoto_2019}. 
\red{The pulsar in the Crab Nebula, known as} the Crab \red{Pulsar} (PSR~B0531+21), was \red{initially} discovered by its GRPs \cite{1968Sci...162.1481S}. Regular periodic emission from the Crab Pulsar \red{occurs }
from low frequency radio to high-energy gamma rays. At 2 GHz (S-band), the GRP emission \red{occurs at} two \red{places}: the main pulse (MP), and the inter-pulse (IP), with a separation of ~0.4 cycles in phase.
\red{Sporadic} GRPs occur both at the MP and IP \red{of} the average  radio pulse, \red{with each} individual GRP \red{lasting} for a much narrower interval ($\sim 3\times 10^{-4}$ in phase) \red{than the regular pulse}. 

\red{Previous studies have searched} for a correlation between radio giant pulses and higher energy emission 
[table \ref{tab:table_previous_observations} and the supplementary materials \cite{Supplement}]. 
The radio flux of GRPs \red{is} 2--3 orders of magnitude \red{higher than} regular pulses\red{; such a large enhancement does not occur at other wavelengths.} \red{An} enhancement \red{of} $\sim$3\% \red{(}$7.8\sigma$ significance) \red{is known} in the optical band (650--700~nm) \cite{2003Sci...301..493S} and \red{independently confirmed} \cite{2013ApJ...779L..12S}. X-ray and gamma-ray  observations have \red{not} detected any statistically significant correlations. 

\red{We searched for} enhancement  in X-rays during GRPs from the Crab Pulsar, using the {\it Neutron star Interior Composition Explorer} ({\it NICER}) X-ray observatory, mounted on the International Space Station \cite{2016SPIE.9905E..1HG}. \red{{\it NICER} has an} effective \red{collecting} area \red{of} 1,900~cm$^{2}$ at 1.5~keV, high time resolution ($<$100~ns), and flexible scheduling. Since launch in 2017, we have monitored the Crab \red{Pulsar} with {\it NICER} for  calibration and scientific purposes. The total average count rate  of the Crab \red{Pulsar} and \red{Nebula} is $1.1\times 10^4$~counts~s$^{-1}$ in the 0.3--10 keV band ($\sim370$~counts per spin cycle), below {\it NICER}'s maximum throughput \red{of} $\sim 3.8\times 10^4$~counts~s$^{-1}$, and thus the data are nearly unaffected by pileup, dead-time, and data transfer losses. We \red{also observed the Crab Pulsar with} two radio telescopes in Japan: the 34~m radio telescope of the Kashima space technology center \cite{2016PASP..128h4502T} and the 64~m radio dish of the Usuda deep space center, both operating at \red{2 GHz} S-band 
[tables \ref{tbl:A1}-\ref{tbl:A4} and figs. \ref{figure:A1}-\ref{figure:A8}; details are given in \cite{Supplement}]. 

\begin{figure}
\begin{center}
  \includegraphics[scale=0.87]{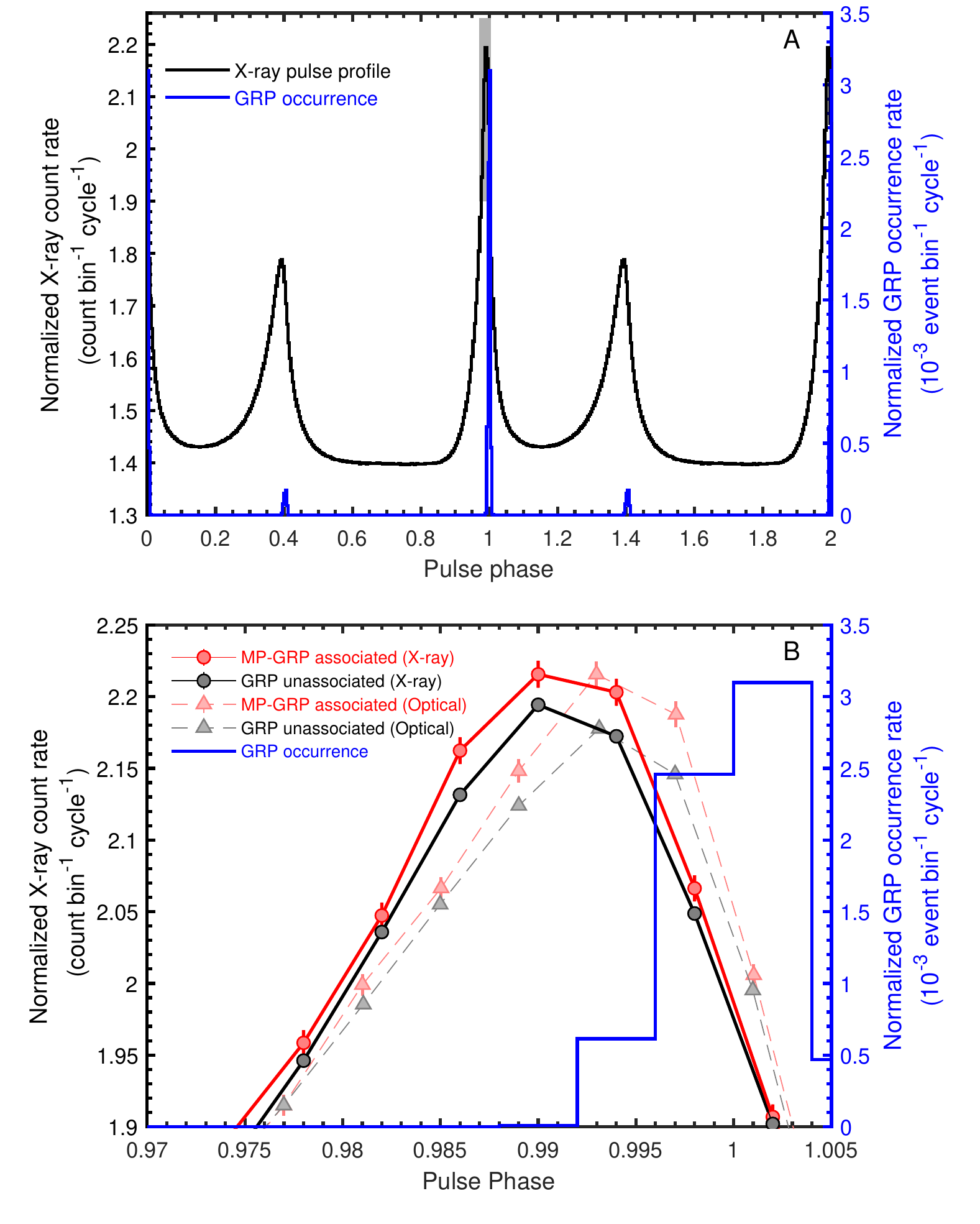}
\vspace{-3mm}  
  \caption{
\textbf{X-ray \red{and optical} pulse \red{profiles} of the Crab \red{Pulsar compared to GRPs}.}
\textbf{(A)} The 0.3--10.0 keV profile (black histogram) \red{observed} with {\it NICER} in 2017--2019 (left axis). The profile is generated with 250 phase bins per spin period, includes the contribution from the Crab \red{Nebula}, and is normalized by the total number of pulsar spin cycles. Two pulse cycles are shown for clarity. The \red{phase} distribution of \red{GRPs} is shown in blue, \red{as measured} from the 2.2--2.3~GHz radio data \red{from} the Usuda and Kashima observatories (right axis). \textbf{(B)} \red{A zoomed view of gray-shaded area of panel A.}
Black and red points \red{connected} with solid lines \red{show} the X-ray profiles without and with GRP association, respectively, with \red{error bars indicating the} 1~$\sigma$ statistical \red{uncertainties} (error bars of the black and gray points are too small to be visible). The \red{blue histogram shows the} GRP-occurrence distribution (identical \red{to} panel A). The \red{faint} dashed lines (black and red) show the optical profiles without and with GRP association, respectively, normalized by an arbitrary scaling \cite{2013ApJ...779L..12S}.  
}
\label{fig:pulse_profile}
\end{center}
\end{figure}

In 2017--2019, we coordinated 15 {\it NICER} observations concurrently with either the Usuda or Kashima observatories [tables \ref{tbl:A1}, \ref{tab:nicer_log} and \cite{Supplement}].
We extracted a total of 126 ks of exposure with simultaneous radio and X-ray coverage \cite{Supplement}. The arrival time of each X-ray photon was converted to  barycentric dynamical time (TDB).
Figure~\ref{fig:pulse_profile}A shows the measured X-ray pulse profile \red{in bins of} 132~\red{\textmu}s. The X-ray MP peak  precedes the average radio profile by \red{the rotation phase} $\Delta \phi \simeq 0.01$ ($\sim$300~\red{\textmu}s), as previously reported \cite{2012A&A...545A.126M}. A constant \red{component from the Crab Nebula} ($1.03\times10^4$~counts~s$^{-1}$) was subtracted \red{before} the following analyses.

During the concurrent coverage, we detected $\sim 2.49\times 10^{4}$ and $\sim 1.75\times 10^{3}$ GRPs at the MP (hereafter referred to as MP-GRPs) and IP (hereafter IP-GRPs) phases at $\phi =$0.9917 \red{to} 1.0083 and $\phi =$1.3944 \red{to} 1.4111, respectively. The occurrence rates of MP-GRPs and IP-GRPs at S-band are 0.67\% (24,851 cycles) and 0.047\% (1,749 cycles), respectively, of the observed 3,731,830 pulsar rotations. Figure~\ref{fig:pulse_profile} shows  the phase \red{distribution of the} GRPs. We defined our GRP samples as those pulses with signal-to-noise ratio exceeding 5.0$\sigma$, which corresponds to a fluence of $\gtrsim 10^3$~Jy~\red{\textmu}s [fig.~\ref{figure:A8} \cite{Supplement}].
The occurrence phases and fractions of the MP- and IP-GRPs are consistent with past measurements \cite{2016ApJ...832..212M,2018PASJ...70...15H}.

\red{We combined} the X-ray photons in \red{three bins, corresponding to} pulsar rotation cycles where MP-GRPs, IP-GRPs, \red{or} neither \red{occurred}. These are hereafter referred to as MP-GRP-associated, IP-GRP-associated and non-GRP-associated X-ray events, respectively. Figure~\ref{fig:pulse_profile}B compares the MP-GRP-associated X-ray profile with the non-GRP-associated profile. The MP-GRP-associated X-ray profile shows an enhancement around the phase of the MP, with similar characteristics to that of \red{the previously} reported optical enhancement \cite{2003Sci...301..493S,2013ApJ...779L..12S}.
Within the pulse phase interval $\phi=$0.985--0.997 [the same width as the optical measurement \cite{2013ApJ...779L..12S}, taking into account the \red{observed} phase shift between the X-ray and optical bands \cite{2014RPPh...77f6901B}], the MP-GRP-associated X-ray profile shows an  enhancement by 3.8$\pm$0.7\% \red{over the non-GRP-associated profile.} 
We \red{performed} the same \red{analysis} for IP-GRP-associated X-rays, but did not \red{find} statistically significant results. We derived a 3$\sigma$ upper limit of the enhancement as 10\% at $\phi=$1.378-1.402 rotational phase \cite{Supplement}. Hereafter we focus on the MP-GRP-associated case.  

\begin{figure}
\begin{center}
 \includegraphics[scale=1.0]{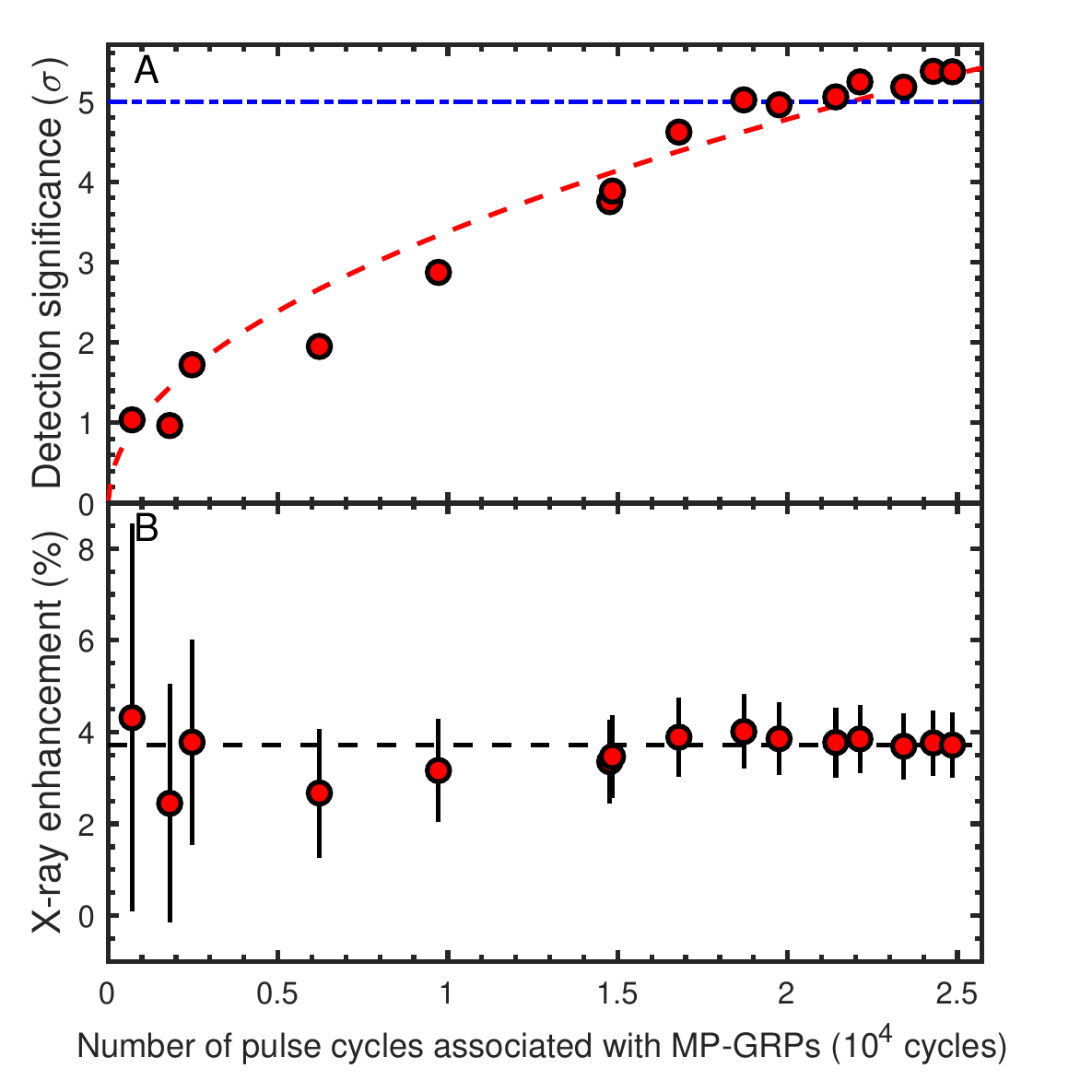}
  \caption{
\textbf{Growth curves of the detection significance and X-ray enhancement.} 
(A) Detection significance as a function of the accumulated number of  rotation cycles \red{observed to be} associated with MP-GRPs. Data accumulation is performed chronologically. Each data point represents all the {\it NICER} data up to \red{that epoch.} The horizontal blue dot-dashed line is the 5$\sigma$ detection significance level and the red dashed curve is the best-\red{fitting model: the} square-root of the cycle number. (B) The same as panel A but for the degree of X-ray enhancement. The horizontal dashed line shows the 3.8\% enhancement ratio, derived from the entire data set. The error bars are statistical 1$\sigma$ \red{uncertainties}.
  } 
  \label{fig:growth_curve}
\end{center}  
\end{figure}

To evaluate the statistical significance of this enhancement, we generated synthetic X-ray samples that have no correlation with MP-GRPs, taking into account the \red{look elsewhere} effect \cite{2010EPJC...70..525G}. We randomly selected  X-ray events with the same number of cycles as the MP-associated ones (24,826 cycles) from the  non-GRP-associated sample. We repeatedly generated 1,000 synthetic control samples and made a histogram of simulated enhancements \cite{Supplement}.
This histogram follows the Gaussian distribution with a mean and standard deviation of $-0.02$\% and 0.70\%, respectively. Thus, the  significance of the measured enhancement \red{is 3.8\% / 0.70\% $=$} 5.4$\sigma$. \red{Figure~\ref{fig:growth_curve} shows} the growth curves of detection significance and the X-ray enhancement rate as a function of the accumulated numbers of MP-GRP-associated cycles. The curves show a monotonic increase of the significance that follows  the square-root of the number of the MP-GRP-associated cycles and a consistent X-ray enhancement rate. This detection is also confirmed by \red{a} lag analysis \cite{Supplement}. 
\red{We did not detect any spectral changes at the MP between the GRP-associated and the non-GRP-associated profiles except for the normalization increase corresponding to the enhancement \cite{Supplement}.}

\begin{figure}
\begin{center}
  \includegraphics[scale=2.0]{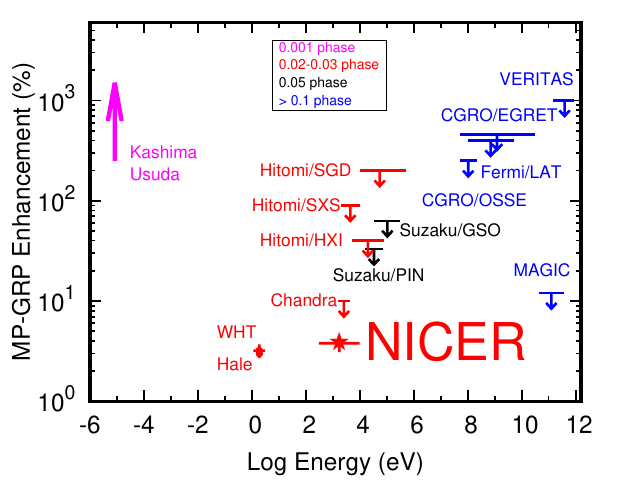}
  \caption{
  \textbf{Enhancement rates with  MP-GRPs as a function of  photon energy}. Optical (WHT and Hale telescope) and {\it NICER} data points are detections. \red{All other data points are upper limits in the}
  X-ray and gamma-ray bands indicated by arrows (\red{values and references are listed in Table \ref{tab:table_previous_observations} \cite{Supplement}}. 
  \red{Magenta,} red, blue, and black colors represent different \red{analysed}
  phase ranges of \red{0.001,} 0.02--0.03, 0.05 and $>0.1$, respectively.
  \red{For our observed radio enhancements with Kashima and Usuda, the detection threshold is shown in magenta arrow. 
See \cite{Supplement} for abbreviations in the figure.} }  
  \label{fig:SED}
\end{center}  
\end{figure}

The total number of  X-ray photons concurrent with radio observations in our analysis is $1.4\times10^9$ counts\red{,} about three orders of magnitude larger than those in  past X-ray studies \red{[e.g.,} \cite{2012ApJ...749...24B,2018PASJ...70...15H}]. Figure~\ref{fig:SED} compares our detection of the X-ray enhancement of MP-GRPs with previous \red{multiwavelength} studies [table \ref{tab:table_previous_observations} \cite{Supplement}].
\red{Our detection of a} 3.8\% X-ray enhancement is consistent with the upper limits  of $\sim$10\%  obtained  in \red{previous} studies and \red{similar} to the  measured \red{3.2\%$\pm0.5$\%} optical enhancement \red{\cite{2013ApJ...779L..12S}}. 
The pulse phase where this X-ray excess appears, $\phi$=0.985--0.997, is also consistent with the reported enhancement phase ($\phi$=0.987--0.999) at optical wavelengths \cite{2013ApJ...779L..12S}. 
This implies that the MP-GRP-associated higher-energy component extends from optical to X-rays without change of its pulse phase or a spectral  \red{cutoff compared} to the average regular pulse.
The X-ray flux of the regular pulsed emission ($\sim4.43\times10^{-9}$~ ergs~s$^{-1}$~cm$^{-2}$ in 0.3--10 keV) is $\sim$1000 and $\sim 10^7$ times higher than those of the optical [$\sim4.6\times10^{-12}$~ ergs~s$^{-1}$~cm$^{-2}$ at 5,500 \AA, \cite{2013ApJS..208...17A}] and regular radio pulses [$\sim1.7\times10^{-16}$ ergs~s$^{-1}$~cm$^{-2}$ at 2 GHz, \cite{2016A&A...591A.134B}], respectively. 
Assuming the same enhancement rate ($\sim$4\%) in both the optical and X-ray bands \red{implies} 1--2 order magnitude higher total energy (\red{both} flux and fluence) emitted from GRP-associated events than the value derived from the radio and optical data only. 
If the X-ray enhancement derived by averaging over $\sim300$~\red{\textmu} s (Figure~\ref{fig:pulse_profile}) consists of multiple short pulses similar to GRP pulses (a typical duration of $\sim 15$~\red{\textmu}s for each \red{GRP}, evaluated as fluence divided by the individual peak flux), the \red{peak} X-ray flux \red{could} be much higher ($\sim20$ times) than the averaged enhancement flux. 

\red{These results constrain} the GRP emission mechanism. The same degree of enhancements ($\sim$4\%) between the optical and X-rays indicates that the GRP-associated high-energy radiation has the same \red{spectral energy distribution} as that of regular pulses. Thus, the spectral energy distribution of GRP-emitting particles is similar to those of particles emitting regular pulses resulting from particle acceleration in the \red{pulsar} magnetosphere or \red{a thin corrugated plasma flow at the equatorial plane (}current sheet\red{)}. The X-ray \red{emission associated with} GRPs \red{implies} that the radio emission efficiency is $\lesssim1\%$, consistent with the expectation from a magnetic reconnection model \cite{2019ApJ...876L...6P}.

\red{A proposed model of} GRP high-energy radiation \red{invokes} a temporal increase in particle number \red{density} in the emitting region \red{\cite{2007MNRAS.381.1190L}}. The \red{difference in} enhancement between the radio (several orders of magnitude) and optical/X-ray bands ($\sim 4$\%) is \red{then} attributed to \red{incoherent (X-ray and optical) emissions are proportional to the particle number, whilst coherent (radio) emission is proportional to the particle number squared} [Supplementary Text in \cite{Supplement}].
Other \red{proposed} mechanisms are emission from high-energy particles in the \red{plasma blobs (}plasmoids\red{)} generated via magnetic reconnection \cite{2019ApJ...876L...6P} and from the resonant absorption of radio photons by X-ray emitting particles \cite{2008ApJ...680.1378H} \cite{Supplement}. 

Bright GRPs from young and energetic pulsars or magnetars \red{have been proposed as} low-energy \red{analogues} of FRBs \cite{2016MNRAS.457..232C} \red{but this proposal has been disputed} \cite{2017ApJ...838L..13L}. 
\red{The proposal relies on the unknown} GRP radio emission efficiency $\eta$ relative to the spin-down luminosity. Even in the case of an extremely high efficiency ($\eta\sim1$), the spin-down timescale for FRB sources \red{are} shorter than $100$ yrs \cite{2017ApJ...838L..13L,2017PASJ...69L...9K}. If FRBs \red{are accompanied by} X-ray emission \red{increases} similar to Crab GRPs, the spin-down rate is enhanced by a factor of $1/\eta$ with $\eta \ll 1$. \red{This would cause rapid radio flux decay,} which is inconsistent with observations of \red{such as the repeating FRB~121102 \cite{2021MNRAS.500..448C}.
Our results therefore disfavour the proposed connection between GRPs and the repeating FRBs.}

\newpage

\input{scifile_Main.bbl}


\section*{Acknowledgments}
{\it NICER} analysis software and data calibration is
provided by the NASA {\it NICER} mission and the Astrophysics
Explorers Program. We thank Y. Terada and N. Kawai for suggestions on our analysis and manuscript.  We thank the Usuda 64-m antenna operation support team in Space Tracking and Communication Center and ISAS, JAXA, and also E.~Kawai and S.~Hasegawa of the Space Time Standard Laboratory of the Kashima Space Technology Center of the National Institute of Information and Communications Technology (NICT) for supporting observations with the Kashima 34-m antenna.
\textbf{Funding:} T.E.. T.T., H.M., K.A., M.H., S.S., S.J.T., and S.K., are supported by JSPS/MEXT KAKENHI grant numbers 15H00845, 15K05069,16H02198, 17H01116, 17K18270, 17K18776, 18H01245, 18H01246, and 18H04584, 19K14712. T.E. acknowledges Hakubi projects of Kyoto University and RIKEN. W.C.G.H. acknowledges
support through grant 80NSSC20K0278 from NASA. 
C.H. was supported as the JSPS International Research Fellow (ID: P18318) and by the the Ministry of Science and Technology in Taiwan through grant MOST 109-2112-M-018-009-MY3. 
C.M. is supported by an appointment to the NASA Postdoctoral Program at the Marshall Space Flight Center, administered by Universities Space Research Association under contract with NASA.
H.T. was supported by the discretionary expenses of the director of Institute of Space and Astronautical Science (ISAS), JAXA. W.A.M carried out research at the Jet Propulsion Laboratory, California Institute of Technology, under a contract with the NASA.
\textbf{Authors contributions:} 
T.E. and T.T. led this radio-X-ray collaboration. T.T. and C.H. led the radio and X-ray timing analyses, respectively. S.G. performed the X-ray spectral analyses. S.K. led the theoretical discussion. P.S.R., T.O., Z.A., K.C.G., C.B.M., Y.S., S.K., S.B., and R.F. were responsible for {\it NICER} detector development, timing calibration, and observation planning. C.M., W.C.G.H., A.K.H., Z.W., W.A.M., T.G., G.K.J., K.A., S.S., N.L. and S.J.T. contributed the theoretical interpretation. 
Y.M., H.T., K.T., M.S., Y.Y., H.M., F.T., T.A., M.T., M.H., O.K., and T.O. were responsible for the radio observations and associated data analyses. N.L. contributed to summarize the previous studies. 
\textbf{Competing Interests:} 
We declare no competing interests. 
\textbf{Data and materials availability:}
The {\it NICER} X-ray data and analysis softwares are available from the NASA HEASARC archive (\url{https://heasarc.gsfc.nasa.gov/}) and HEASoft (\url{https://heasarc.gsfc.nasa.gov/docs/software/heasoft/}), respectively. The observation IDs (ObsIDs) are listed in \cite{Supplement}. The GRP event list and analysis source codes to reproduce the results of this article are available on Zenodo \cite{Data}. 
\clearpage

\clearpage

\clearpage

\clearpage
{\bf Supplementary Materials}\\
\begin{itemize}
    \item Materials and Methods
    \item Supplementary Text
    \item Figures S1 to S20
    \item Tables S1 to S7
    \item References (33--78)
    \item Movie S1
\end{itemize}

\begin{figure}
\begin{center}
\vspace*{-3cm}
\centerline{\includegraphics[scale=1.0]{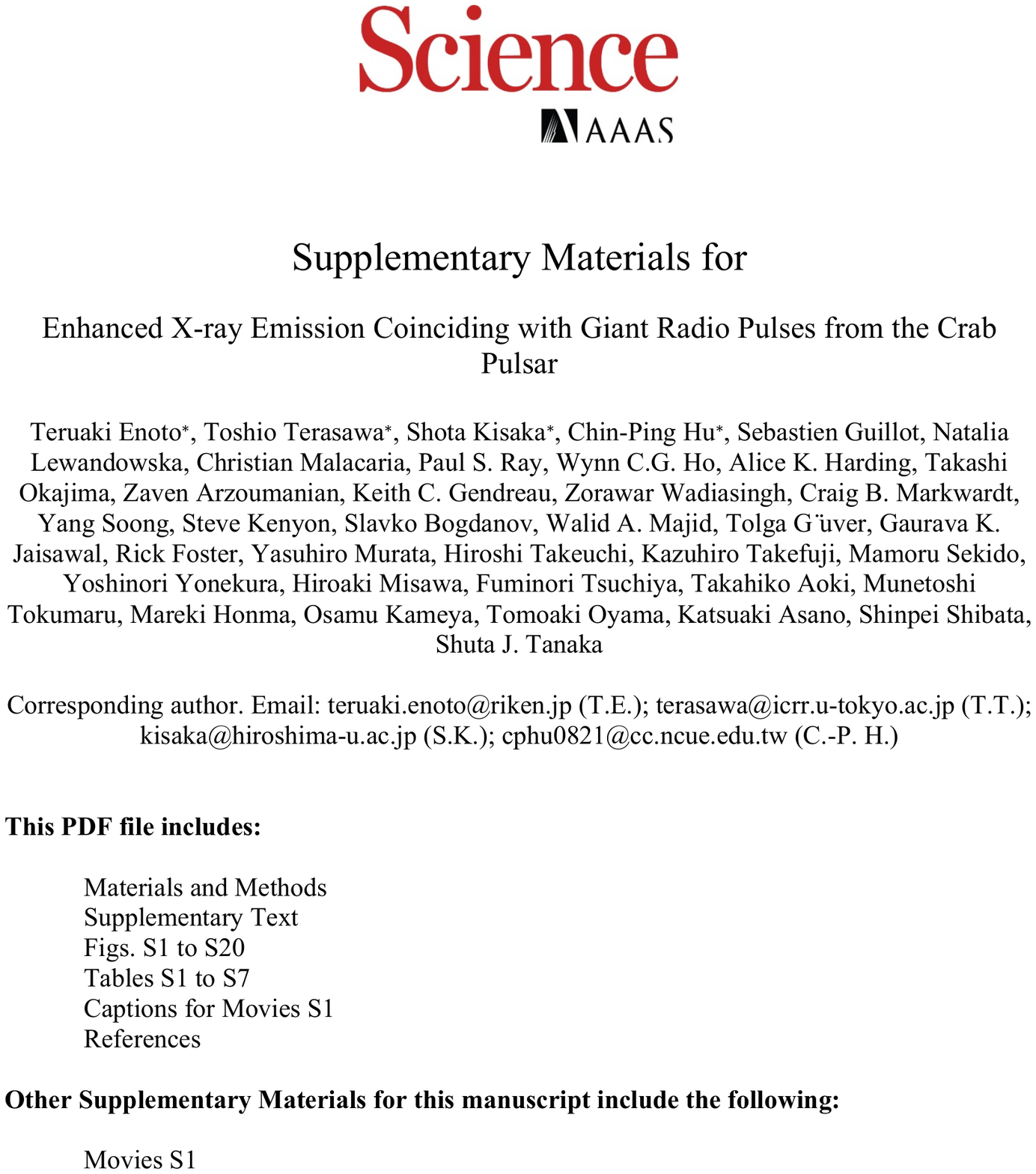}}
\end{center}  
\end{figure}

\clearpage
\section*{Materials and Methods}

\renewcommand\thetable{S\arabic{table}} 
\setcounter{table}{0} 

\renewcommand\thefigure{S\arabic{figure}}
\setcounter{figure}{0}

\renewcommand\theequation{S\arabic{equation}}
\setcounter{equation}{0}

\input{suppl_01_previous_studies_v2}

\clearpage

\input{suppl_02_radio_ver0204_v2}

\clearpage
\input{suppl_03_xray_v2}

\clearpage
\section*{Supplementary Text}
\input{suppl_04_theorical_v2}

\newpage

\begin{scriptsize}
\begin{table}
  \begin{center}
    \caption{\textbf{Previous optical, X-ray, and gamma-ray searches for correlations with the Crab Pulsar GRPs.}} \vspace{3mm}
    \label{tab:table_previous_observations}
    \begin{tabular}{lllll} 
    \hline \hline
    Band & Observatory & MP & IP & Ref. \\
    \hline
Optical    & William Herschel Telescope  & 3\% ($7.8\sigma$) & $<$2.5\% ($1\sigma$) & \cite{2003Sci...301..493S} \\ 
(600-750 nm) & (WHT) & & & \\ 
Optical    & Hale Telescope &    3.2\% (7.2$\sigma$) & 2.8\%(3.5$\sigma$) & \cite{2013ApJ...779L..12S} \\
(1.1-3.1 eV) & & & & \\
Soft X-ray & NICER   & 3.8\% (5.4$\sigma$) & $<$10\% (3$\sigma$) &  This work \\
(0.3--10 keV) & & & & \\
Soft X-ray & Chandra    & $<$10\% (2$\sigma$) & $<$30\% (2$\sigma$) & \cite{2012ApJ...749...24B} \\
(1.5--4.5 keV) & & & & \\
X-ray & Hitomi    & $<$25\% (3$\sigma$) &    $<$110\% (3$\sigma$) & \cite{2018PASJ...70...15H} \\
(2--300 keV) & & & & \\
Hard X-ray & Suzaku / HXD-PIN & $<$33\% (2$\sigma$) & $<$88\% (2$\sigma$) & \cite{2015PhDT.......M} \\
(15--75 keV) & & & & \\
Hard X-ray & Suzaku / HXD-GSO & $<$63\% (2$\sigma$) & $<$193\% (2$\sigma$) & \cite{2015PhDT.......M} \\
(35--315 keV) & & & & \\
Hard X-rays & OSSE / CGRO & 
\multicolumn{2}{c}{$<$250\%    (1$\sigma$)} & \cite{1995ApJ...453..433L} \\
(50--220 keV) & & & & \\
Gamma ray & EGRET / CGRO & $<$460\% (3$\sigma$) & -- & \cite{1998ApJ...496..863R} \\
(0.05--30 GeV) & & & & \\
Gamma ray & LAT / Fermi & $<$400\% (2$\sigma$) & $<$1200\% (2$\sigma$) & \cite{2011ApJ...728..110B} \\ 
(0.1-5 GeV) & & & &  \\ 
Gamma ray & LAT / Fermi & -- & -- & \cite{Mickaliger_2012}$^{\ast}$
\\ 
(0.1-100 GeV) & & & &  \\
VHE gamma ray & VERITAS & -- & $<$500--1000\% (2$\sigma$) & \cite{2012ApJ...760..136A} \\
($>$150 GeV) & & & &  \\
VHE gamma ray & MAGIC & \multicolumn{2}{c}{$<$12--2900\% (2$\sigma$)} & \cite{2020A&A...634A..25M} \\
($>$60 GeV) & & & &  \\
\hline
\end{tabular}
\begin{flushleft}
($\ast$)  That study did not provide upper limits of the flux normalized to the pulsed gamma-rays of the Crab Pulsar due to insufficient detected gamma-ray photons. 
\end{flushleft}
  \end{center}
\end{table}
\end{scriptsize}

\begin{scriptsize}
\begin{table}
\begin{center}
\caption{\textbf{Basic parameters of radio observatories.}
A radio ID of the observatories (U or K) refers to Tables \ref{tbl:A2} and \ref{tbl:A4}.
The system equivalent flux density (SEFD, the fourth column) has an uncertainty of $\sim$10\%.
To derive the effective total band width $\Delta \nu$ (the rightmost column), 
we take into account the gain reduction at the channel edges.
}
\label{tbl:A1}
$$
\begin{array}{cccccc}
\hline \hline
{\rm Observatory} &{\rm dish~size}& {\rm Geographical} &  {\rm SEFD}  & {\rm frequency~coverage}  &{\rm effective~total~band}      \\
({\rm Radio~ID}) &{\rm (m)}      & {\rm position}     &  {\rm (Jy)}   & {\rm (MHz)}               &{\rm width},~ \Delta \nu~{\rm (MHz)}        \\
\hline 
{\rm Usuda~(U)}        &64             & 138^{\rm o}22'{\rm E} & 105 & {\rm ch0\mathchar`-3}: 2194\mathchar`-2322                     & 120  \\
                       &               &  36^{\rm o}08'{\rm N} &     &                               &      \\
\hline
{\rm  Kashima~(K)}     &34             & 140^{\rm o}40'{\rm E} & 476 & ~~{\rm ch0}:~2194\mathchar`-2226         &  90  \\
                       &               &  35^{\rm o}57'{\rm N} &     & {\rm ch2,3}:2258\mathchar`-2322         &      \\
\hline
\end{array}
$$
\end{center}
\end{table}
\end{scriptsize}

\begin{tiny}
\begin{table}
\begin{center}
\caption{\textbf{Spin ephemeris of the Crab Pulsar at each observing session.}
MJD (the second column) stands for Modified Julian day.
Spin frequency $\nu_0$ (the fifth column) and its derivative $\dot{\nu}_0$ (the sixth column)
 are interpolated from the monthly Jodrell Bank Observatory ephemeris \cite{Jodrellbank}.
DM (the seventh column) stands for dispersion measure, 
which is equivalent to the electron column density along the line of sight,
a mixed unit of which, 1 pc cm$^{-3}$, is 3.08568$\times 10^{18}$ cm$^{-2}$.
Delay (the eighth column) is a group delay time of the radio signal at 2,322~MHz from the X-rays, derived from the DM value.
$y_0$ (the rightmost column) is a daily initial spin phase (Spin number, phase, and GRP identification subsection) 
and its uncertainty (in parentheses) at the last digit.
}
\label{tbl:A2}
\scalebox{0.8}{
\begin{tabular}{ccccccccc}
\hline \hline
Session & 
MJD & year/mm/dd & Radio  & $\nu_0$  &   $\dot{\nu}_0$ 
& DM         & Delay  & $y_0({\rm uncertainty})$   \\
 number  &                                             
              &  (day of year) & ID    & (Hz) & $(\times 10^{-15} {\rm s}^{-2})$   & (pc cm$^{-3}$) &  (ms) & \\
\hline 
1  &57974 & 2017/08/09  ~(221) & U  & 29.6396012136 & $-$368706.87 & 56.7655                 &43.684 & $-$0.16208(3) \\ 
2  &58067 & 2017/11/10  ~(314) &  U  & 29.6366534594 & $-$369627.13 & 56.7508                 &43.671 & $-$0.98615(5) \\ 
3  &58117 & 2017/12/30  ~(364) &  U  & 29.6350532496 & $-$369981.74 & 56.7544                 &43.670 & $-$0.93027(7) \\ 
4  &58121 & 2018/01/03  ~(003) &  U,K & 29.6349254479 & $-$369741.20 & 56.7548                 &43.670 & $-$0.60833(3) \\
5  &58190 & 2018/03/13  ~(072) &  K  & 29.6327232229 & $-$369158.00 & 56.7507                 &43.671 & $-$0.31502(3) \\ 
6  &58191 & 2018/03/14  ~(073) &  K  & 29.6326913277 & $-$369157.20 & 56.7507                 &43.671 & $-$0.40626(3) \\ 
7  &58215 & 2018/04/07  ~(097) &  K  & 29.6319259528 & $-$369011.60 & 56.7517                 &43.665 & $-$0.24444(4) \\ 
8  &58430 & 2018/11/08  ~(312) &  U  & 29.6250752831 & $-$368624.30 & 56.7676                 &43.684 & $-$0.53972(6) \\ 
9  &58431 & 2018/11/09  ~(313) &  U  & 29.6250434340 & $-$368623.50 & 56.7706                 &43.687 & $-$0.41147(5) \\ 
10 &58478 & 2018/12/26  ~(360) &  K  & 29.6235466420 & $-$368594.57 & 56.7941                 &43.701 & $-$0.20133(6) \\ 
11 &58479 & 2018/12/27  ~(361) &  K  & 29.6235148058 & $-$368593.78 & 56.7941                 &43.701 & $-$0.14722(4) \\ 
12 &58480 & 2018/12/28  ~(362) &  K  & 29.6234829593 & $-$368592.99 & 56.7942                 &43.701 & $-$0.84398(5) \\ 
13 &58481 & 2018/12/29  ~(363) &  U  & 29.6234511129 & $-$368592.20 & 56.7943                 &43.701 & $-$0.29158(5) \\ 
14 &58533 & 2019/02/19  ~(050) &  K  & 29.6217953644 & $-$368510.56 & 56.7612                 &43.673 & $-$0.72946(5) \\ 
15 &58725 & 2019/08/30  ~(242) &  U  & 29.6156841478 & $-$368412.06 & 56.7431                 &43.667 & $-$0.44901(5) \\ 
\hline                                         
\end{tabular}
}
\end{center}
\end{table}
\end{tiny}

\begin{figure}
\begin{center}
  \includegraphics[width=10cm]{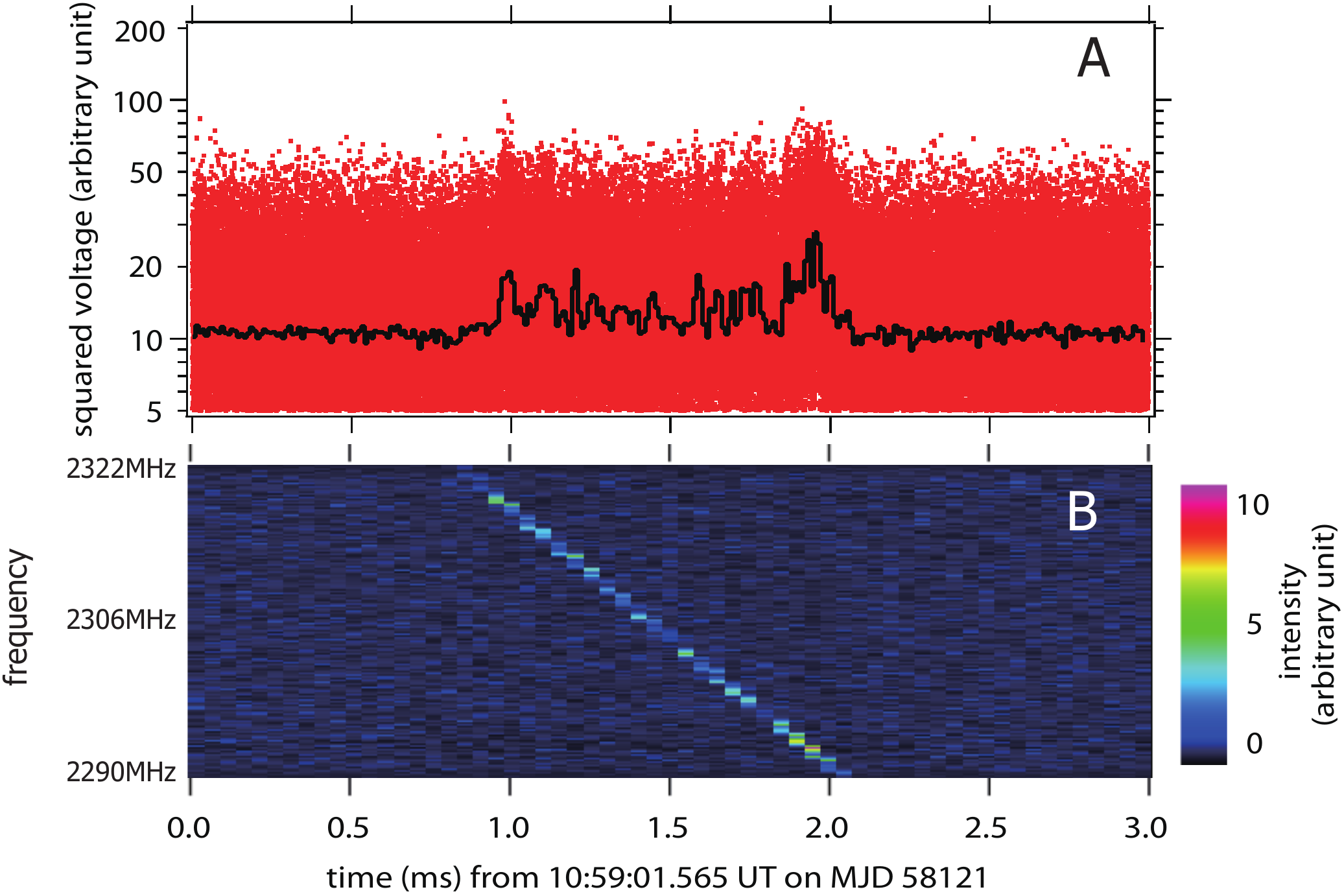}
\caption{
\textbf{An example of MP GRPs, detected  at 
 MJD 58121 (session 4) 10:59:01 
UT (universal time)} in Usuda. 
\textbf{(A)} Red dots  show the squared antenna voltages in channel 3 (before de-dispersion) 
for a 3-ms duration (10:59:01.565--10:59:01.568 UT) with the original time resolution
$\delta t$=15.625 ns, and black line  shows their 10~\textmu s averages. 
\textbf{(B)} Pseudo-color image  of the dynamic spectrum for the same  duration as of (A).}
\label{figure:A1}
\end{center}
\end{figure}

\begin{figure}
\begin{center}
  \includegraphics[width=10cm]{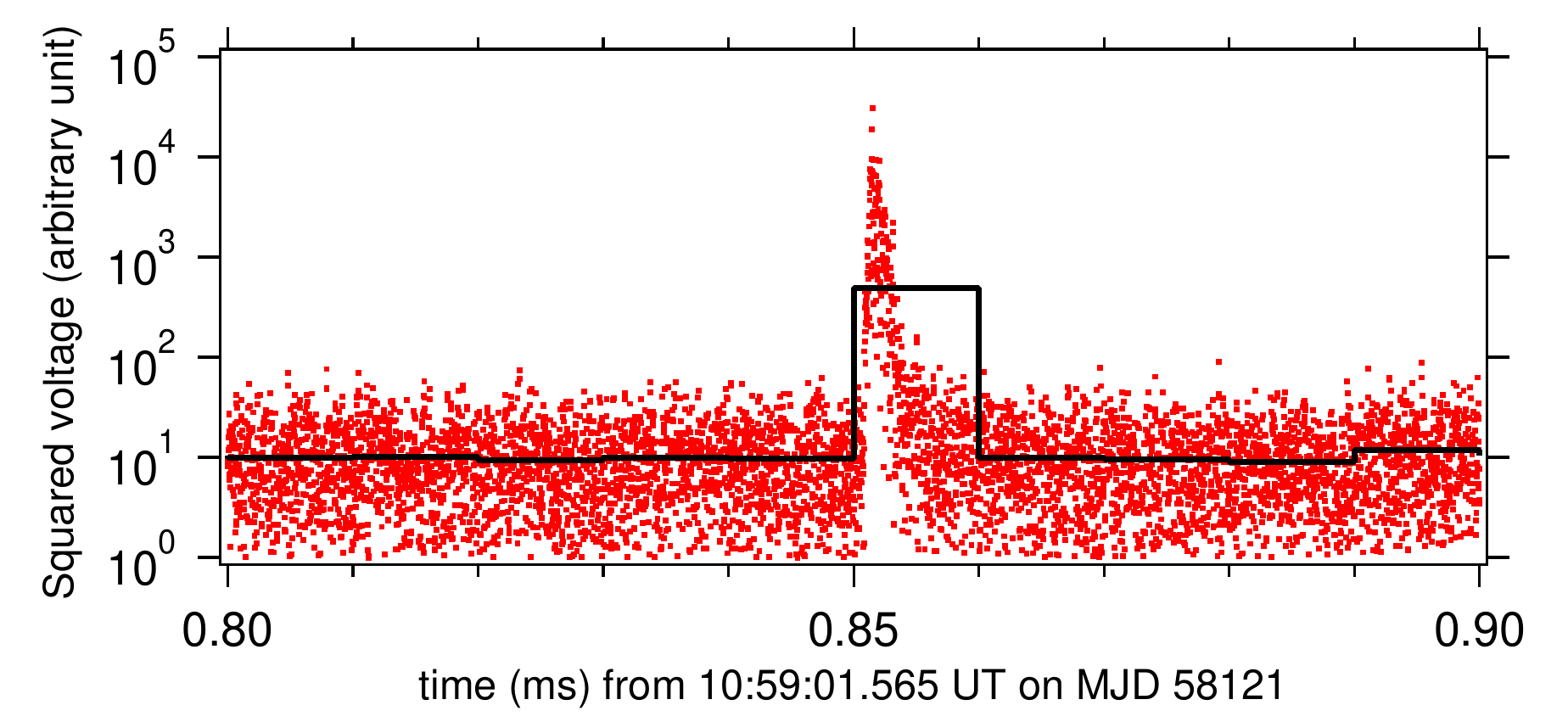}
\caption{
\textbf{A result of the coherent de-dispersion procedure for the same MP GRP 
as Fig.~\ref{figure:A1}A.}
Red dots and a black line respectively show 
$|V^{\rm de\mathchar`-dis}_{\rm ch3} (t)|^2$ 
and $\overline{|V^{\rm de\mathchar`-dis}_{\rm j}(t_k)|^2}$ for a duration of 0.1~ms centered at MJD 58121 10:59:01.56585 UT.
}
\label{figure:A2}
\end{center}
\end{figure}

\begin{figure}
  \begin{center}
   \includegraphics[width=8cm]{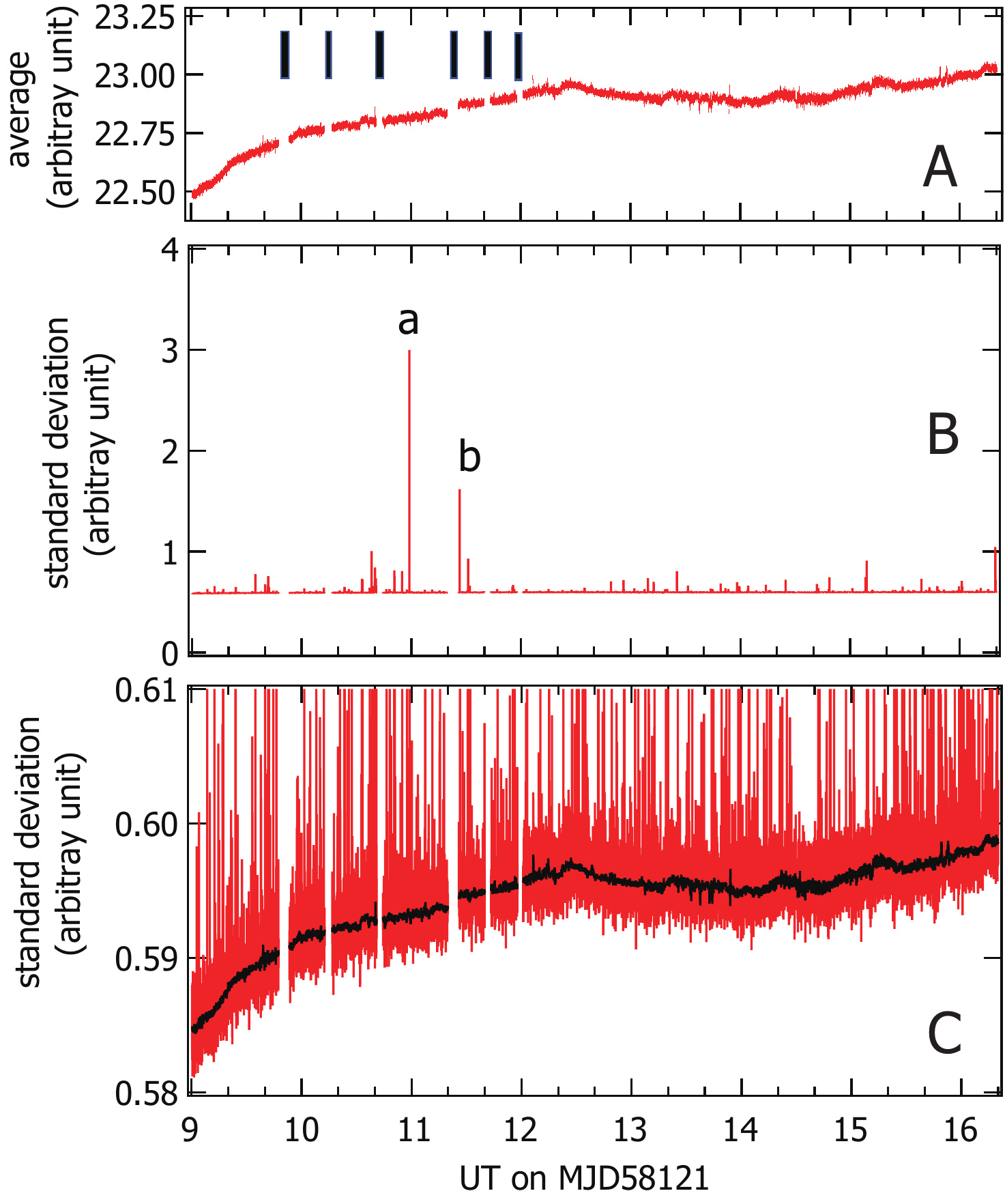}
  \end{center}
 \caption{
\textbf{Averages and standard deviations.}
\textbf{(A)} $\overline{\cal W}(t_N)$ with black blocks showing data gaps caused by the observation schedule.
\textbf{(B)} standard deviations, $\sigma (t_N)$. Spikes marked a and b correspond to MP GRPs (see text).
\textbf{(C)} $\sigma (t_N)$ with an enlarged vertical scale (red line), where $\overline{\sigma} (t_N) \equiv 0.026 \overline{\cal W}(t_N)$ is overlaid (black line).
}
\label{figure:A3}
\end{figure}

\begin{figure}
\begin{center}
  \includegraphics[width=15cm]{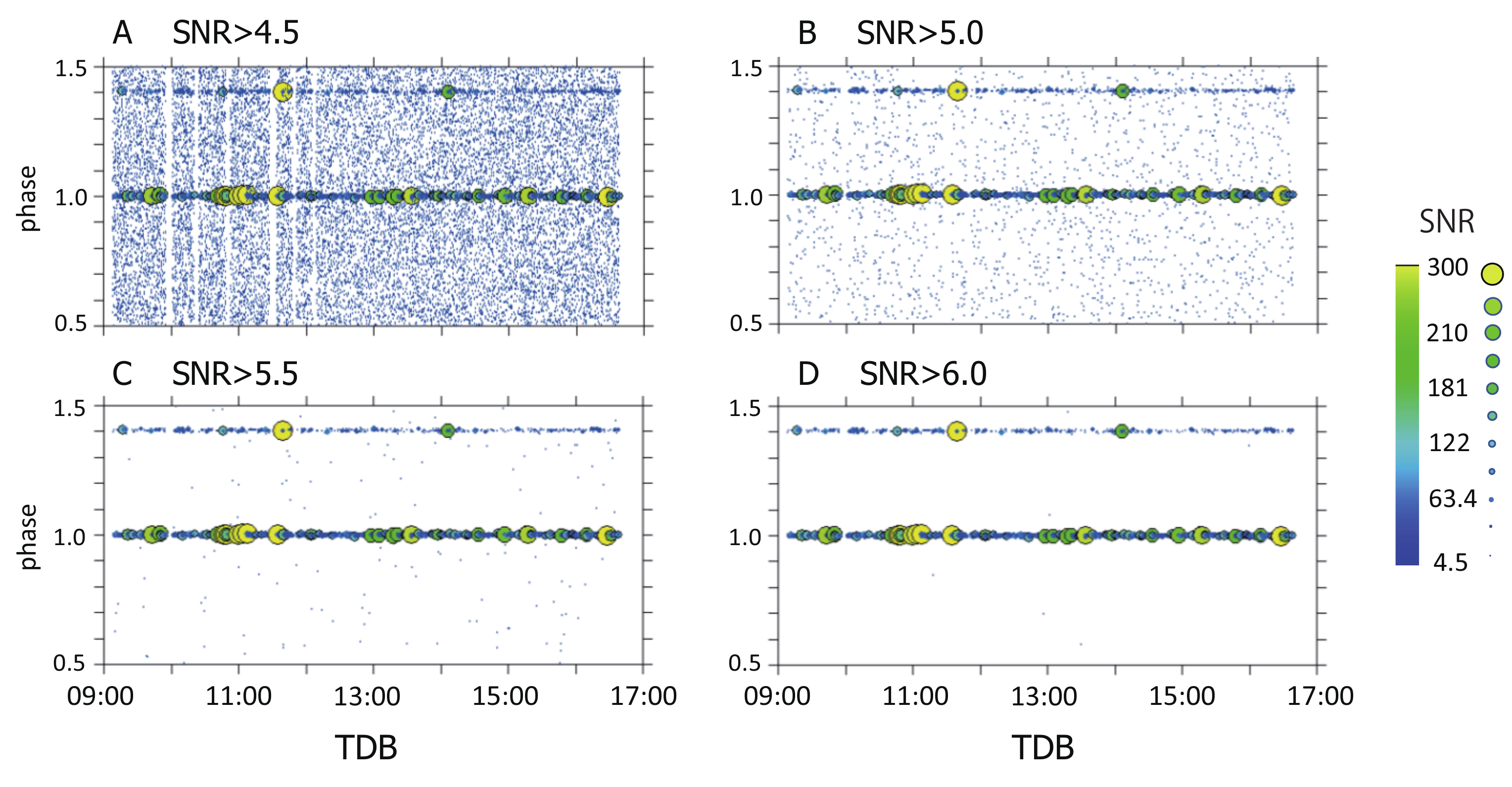}
\caption{
\textbf{Time versus phase diagrams for candidate GRPs at observing session 4.} 
In panels A - D, four different threshold values of the signal to noise ratio, SNR$_{\rm thr}$=4.5, 5.0, 5.5, and 6.0 are adopted.  
Data points for MP GRPs and IP GRPs cluster at around phase=1 and phase=1.406, respectively. 
Homogeneously distributed background points originate in emission from the Crab Nebula, 
and the system thermal noise. 
Data gaps seen in the interval of 09:57--12:10 TDB are caused by the observation schedule, and not by RFI.
}
\label{figure:A4}
\end{center}
\end{figure}

\begin{scriptsize}
\begin{table}
\begin{center}
\caption{\textbf{False GRP rates.}
Numbers of MP-GRP candidates at Usuda and Kashima, $\displaystyle{\sum_{i\in {\rm [U]}}} N^{\rm MP}_{{\rm obs},i}$ and $\displaystyle{\sum_{i\in {\rm [K]}}} N^{\rm MP}_{{\rm obs},i}$, respectively. Those for IP-GRPs are  $\displaystyle{\sum_{i\in {\rm [U]}}} N^{\rm IP}_{{\rm obs},i}$ and $\displaystyle{\sum_{i\in {\rm [K]}}} N^{\rm IP}_{{\rm obs},i}$. The numbers are sums over all the observing sessions and include both candidates and false GRPs numbers.}
\label{tbl:A3}
\begin{tabular}{p{2cm}|rrrr|rrrr}
\hline \hline 
SNR$_{\rm thr}$ & \multicolumn{4}{c|}{MP-GRP} & \multicolumn{4}{c}{IP-GRP} \\ 
for GRP & \multicolumn{2}{c}{GRP Candidates} & \multicolumn{2}{c|}{False rates  } & \multicolumn{2}{c}{GRP Candidates} & \multicolumn{2}{c}{False rates  } \\ 
 & Usuda & Kashima & Usuda & Kashima & Usuda & Kashima & Usuda & Kashima \\ \hline 
4.5 & 43,356 & 53,072 & 8.2\% & 11\% & 5,724 & 8,798 & 62\% & 69\% \\ 
5.0 & 34,931 & 40,940 & 1.1\% & 1.7\% & 2,410 & 3,005 & 17\% & 24\% \\ 
5.5 & 30,392 & 35,162 & 0.11\% & 0.20\% & 1,767 & 2,069 & 1.9\% & 3.4\% \\ 
6.0 & 26,493 & 30,911 & 0.013\% & 0.036\% & 1,524 & 1,747 & 0.23\% & 0.63\% \\ 
\hline 
\end{tabular}
\end{center}
\end{table}
\end{scriptsize}

\begin{figure}
\begin{center}
  \includegraphics[width=15cm]{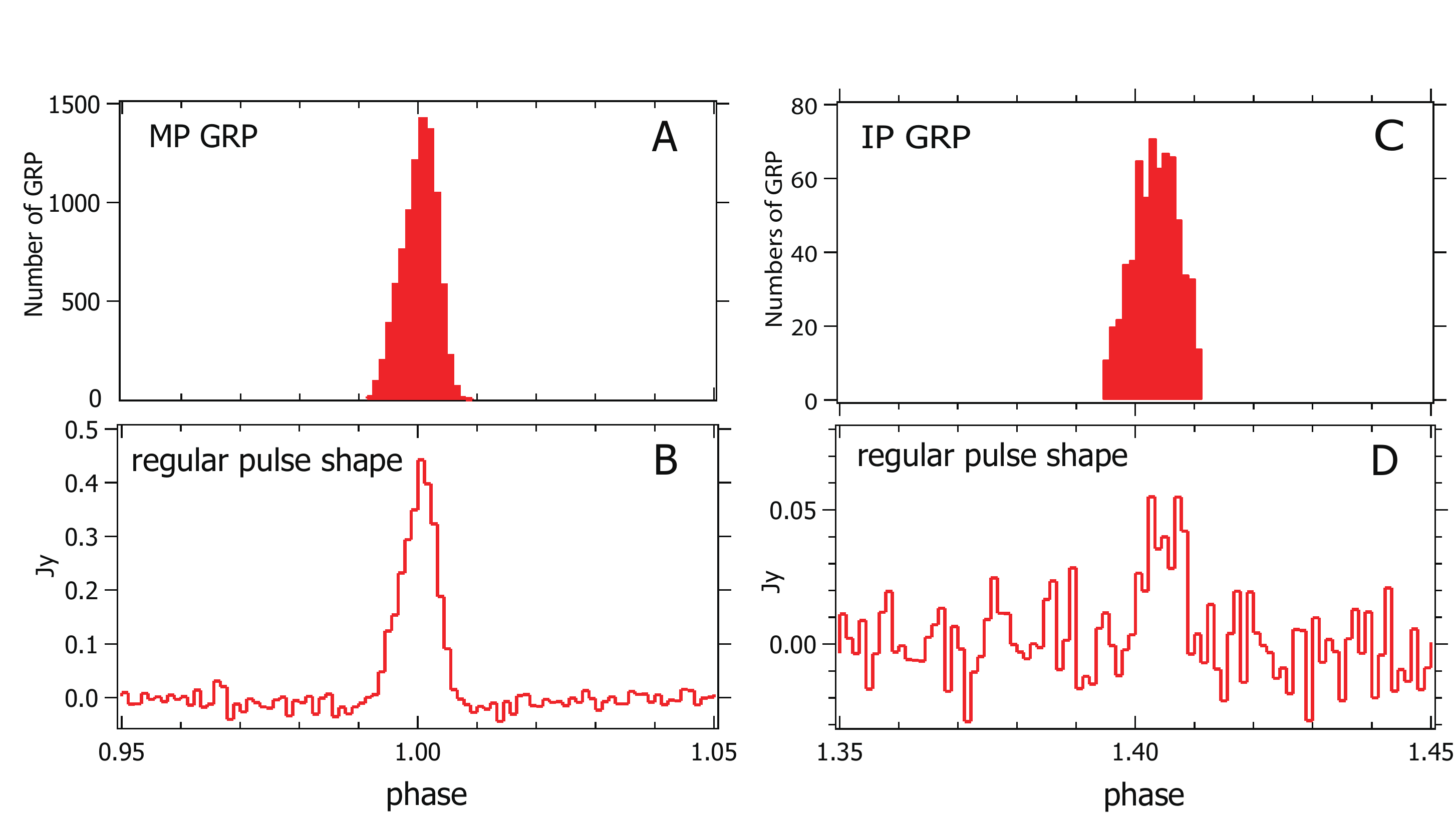}
\caption{\textbf{Number histograms of GRPs [(A) for MP GRPs and (B) for IP GRPs] and regular pulse shapes [(B) for MP phase and (D) for IP phase].} 
Data are from observing session 4. In panels (A) and (C), 
the ordinates show the numbers of GRPs in a bin with a size of a phase width of 1.11$\times 10^{-3}$ equivalent to an angular width of 0.4$^{\rm o}$). 
We represent the ordinates of panels (B) and (D) with the physical unit Jy, utilizing the procedure described in Flux density and fluence subsection.
}
\label{figure:A5}
\end{center}
\end{figure}

\begin{tiny}
\begin{table}
\begin{center}
\vspace{1mm}
\caption{\textbf{Time variations of GRP rates.}
The two rightmost columns shows the MP GRP and IP GRP detection rates.
In the observing session 4, data  of the Kashima observatory are available in addition to those of the Usuda observatory.
Because the Usuda observatory has a higher sensitivity, we use the Usuda data for radio--X-ray correlation analysis.
The Kashima data are used for relative calibration  to the Usuda data. }
\label{tbl:A4}
\begin{tabular}{cccccccc}
\hline \hline
Session &
MJD & Radio & Total~radio   & \multicolumn{2}{c}{Number of GRPs}  & \multicolumn{2}{c}{GRP rate (GRP s$^{-1}$)}    \\  
number  &        & ID  & observation time (ks) & MP     &   IP      & MP     & IP       \\
\hline
1   &  57974     & U     &  25.88 &  8355        &563   &  0.3227 &   0.0218    \\
2   &  58067     & U     &  12.33 &  3892        &265   &  0.3156 &   0.0215    \\
3   &  58117     & U     &   9.05 &  1473        &101   &  0.1627 &   0.0112    \\
4$^*$ &  58121     & U     &  25.44 &  9062        &643   &  0.3563 &   0.0253    \\
5   &  58190     & K     &  37.76 & 10687        &705   &  0.2830 &   0.0187    \\
6   &  58191     & K     &  35.14 & 12908        &896   &  0.3673 &   0.0255    \\
7   &  58215     & K     &  22.76 &  5009        &358   &  0.2201 &   0.0157    \\
8   &  58430     & U     &  10.75 &  2652        &149   &  0.2467 &   0.0139    \\
9   &  58431     & U     &  16.61 &  2893        &197   &  0.1742 &   0.0119    \\
10  &  58478     & K     &  27.59 &  2374        &211   &  0.0860 &   0.0076    \\
11  &  58479     & K     &  39.32 &  4306        &337   &  0.1095 &   0.0086    \\
12  &  58480     & K     &  39.38 &  2698        &282   &  0.0685 &   0.0072    \\
13  &  58481     & U     &  35.90 &  3121        &253   &  0.0869 &   0.0070    \\
14  &  58533     & K     &  21.79 &  2958        &216   &  0.1358 &   0.0099    \\
15  &  58725     & U     &  14.36 &  3483        &239   &  0.2426 &   0.0166    \\
\hline                                                                                                   
 sum &      & U+K    & 374.07 & 75871        & 5415  &  0.2028 &   0.0145    \\
\hline                                                                                                   
partial sum && U     & 150.32 & 34931        & 2410  &  0.2324 &   0.0160    \\
                 &&  K     & 223.74 & 40940        & 3005  &  0.1830 &   0.0134    \\
\hline
\end{tabular}
\end{center}
\end{table}
\end{tiny}

\begin{figure}
\begin{center}
  \includegraphics[width=10cm]{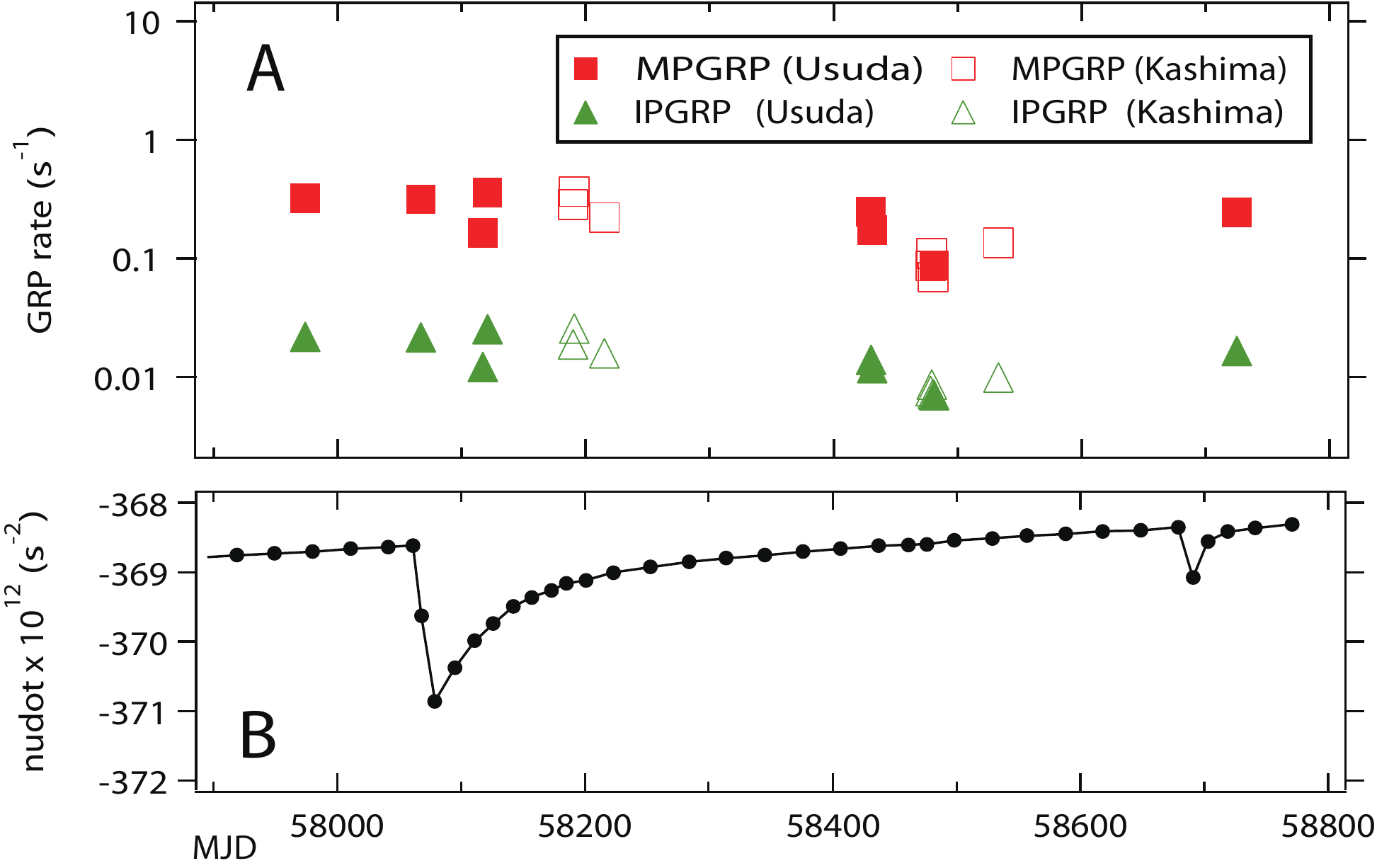}
\caption{
\textbf{Time variations of the GRP rates (A) compared with 
the history of $\dot{\nu}_0$ (B).}
Sharp drops of $\dot{\nu}_0$ in panel (B) on MJD$=$58064 and 58691 
are due to pulsar glitches \cite{2018MNRAS.478.3832S, 2019ATel12957....1S}.
No change in the GRP rate is apparent.
}
\label{figure:A6}
\end{center}
\end{figure}

\begin{figure}
  \begin{center}
   \includegraphics[width=7cm]{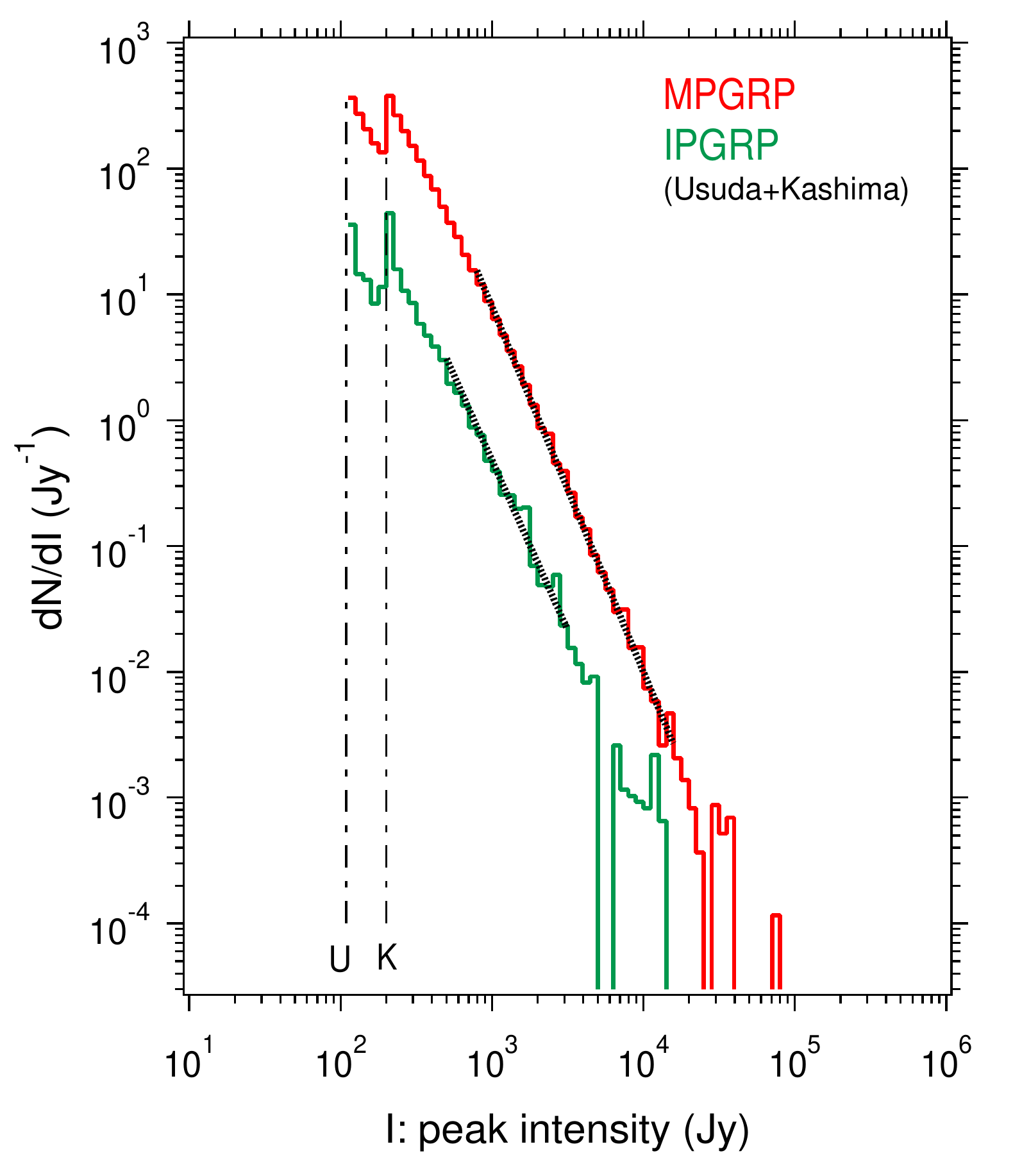}
  \end{center}
 \caption{
\textbf{Differential flux histograms $dN/dI$ of MP GRPs (red) and IP GRPs (green).}
Vertical dashed-and-dotted lines with 'U' and 'K' marks show the threshold values for GRP selection criteria for Usuda and Kashima, respectively.
Sawtooth-like features near the left-side edges of the histograms are artifacts of the different threshold values and not physical.
Results of the power-law fitting (see text) are shown by black dots.
}
\label{figure:A7}
\end{figure}

\begin{figure}
  \begin{center}
   \includegraphics[width=7cm]{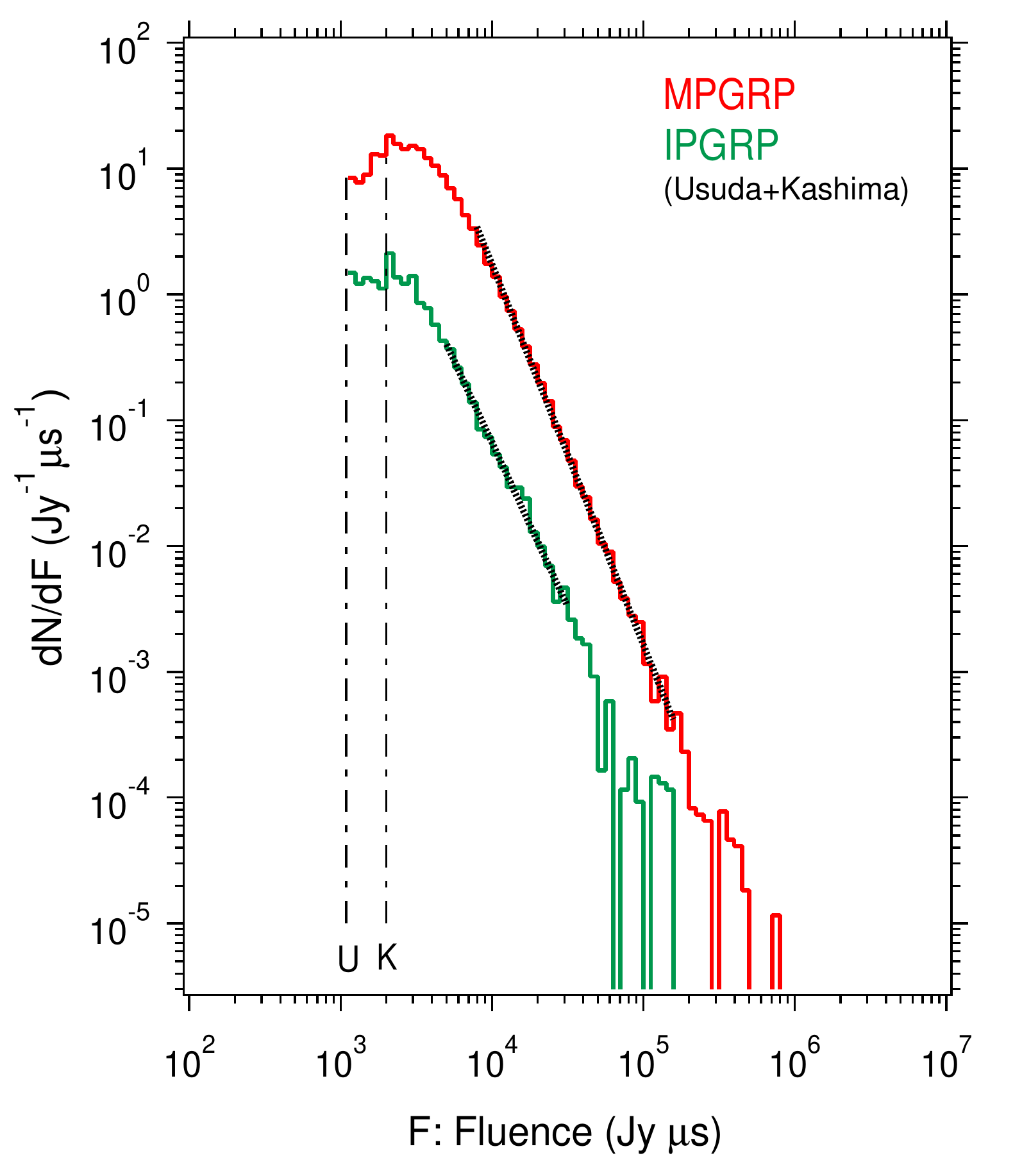}
  \end{center}
 \caption{\textbf{Differential fluence histograms $dN/dF$ drawn with the same format as Fig.~\ref{figure:A7}.} }
\label{figure:A8}
\end{figure}

\begin{figure}
\begin{center}
  \includegraphics[scale=0.6]{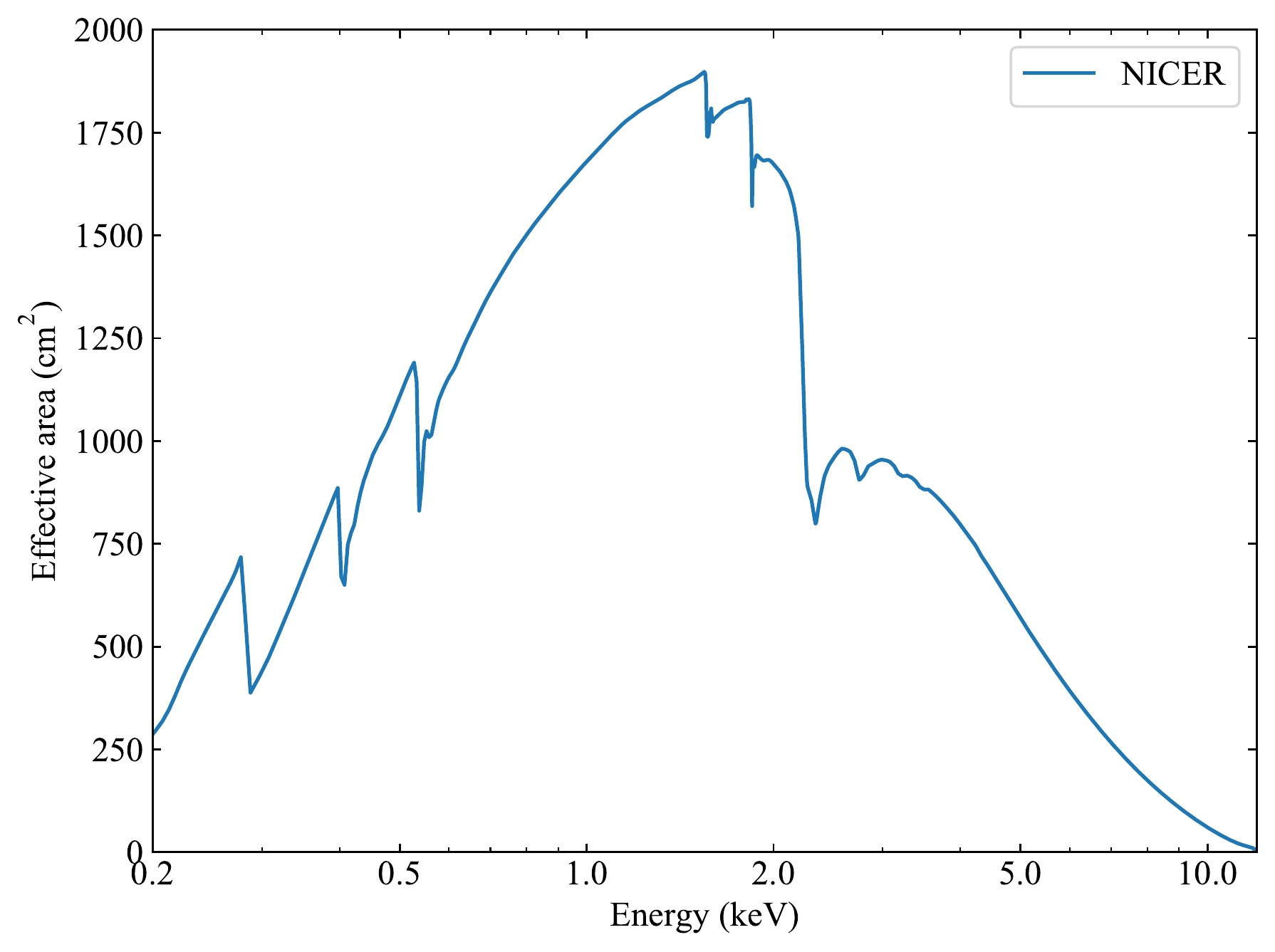}
  \caption{\textbf{Effective area of {\it NICER} \cite{2016SPIE.9905E..1HG}.}}
  \label{fig:effective_area}
\end{center}
\end{figure}

\begin{scriptsize}
\begin{table}
\begin{center}
\caption{\textbf{NICER observations used in  our analysis}. The exposure time is defined as the duration in which the target was simultaneously observed with NICER and radio telescopes. The numbers of the X-ray photons (in 0.3--10 keV), MP-GRPs, and IP-GRPs are  for those detected during the exposure time. }
\label{tab:nicer_log}
\begin{tabular}{rlllrlrr} 
\hline \hline
Session & ObsID & MJD & year/mm/dd & Exposure & Number of  & \multicolumn{2}{c}{Number of GRPs}\\ 
 ID & & & & (ks) & X-ray photons & MP & IP \\
\hline
1 & 1013010104 & 57974 & 2017/08/09 & 2.07 & $2.25\times10^7$ & 709 & 34 \\ 
2 & 1013010110 & 58067 & 2017/11/10 & 3.56 & $3.91\times10^7$ & 1108 & 68 \\ 
3 & 1013010122 & 58117 & 2017/12/30 & 3.81 & $4.20\times10^7$ & 657 & 50 \\ 
4 & 1013010123 & 58121 & 2018/01/03 & 10.27 & $1.13\times10^8$ & 3746 & 271 \\ 
5 & 1013010125 & 58190 & 2018/03/13 & 12.64 & $1.40\times10^8$ & 3504 & 222 \\ 
6 & 1013010126 & 58191 & 2018/03/14 & 13.61 & $1.51\times10^8$ & 5042 & 341 \\ 
7 & 1013010131 & 58215 & 2018/04/07 & 0.41 & $4.52\times10^6$ & 80 & 3 \\
8 & 1013010143 & 58430 & 2018/11/08 & 7.81 & $8.57\times10^7$ & 1955 & 112 \\
9 & 1013010144 & 58431 & 2018/11/09 & 11.26 & $1.24\times10^8$ & 1910 & 149 \\
10 & 1013010147 & 58478 & 2018/12/26 & 12.11 & $1.33\times10^8$ & 1039 & 113 \\
11 & 1013010148 & 58479 & 2018/12/27 & 14.84 & $1.63\times10^8$ & 1668 & 117\\
12 & 1013010149 & 58480 & 2018/12/28 & 10.18 & $1.12\times10^8$ & 709 & 68\\
13 & 1013010150 & 58481 & 2018/12/29 & 14.55 & $1.60\times10^8$ & 1290 & 96\\
14 & 1013010152 & 58533 & 2019/02/19 & 6.51 & $7.15\times10^7$ & 870 & 68\\
15 & 2013010106 & 58725 & 2019/08/30 & 2.36 & $2.57\times10^7$ & 564 & 37 \\
\hline
\end{tabular}
\end{center}
\end{table}
\end{scriptsize}

\begin{figure}
\begin{center}
  \includegraphics[scale=0.6]{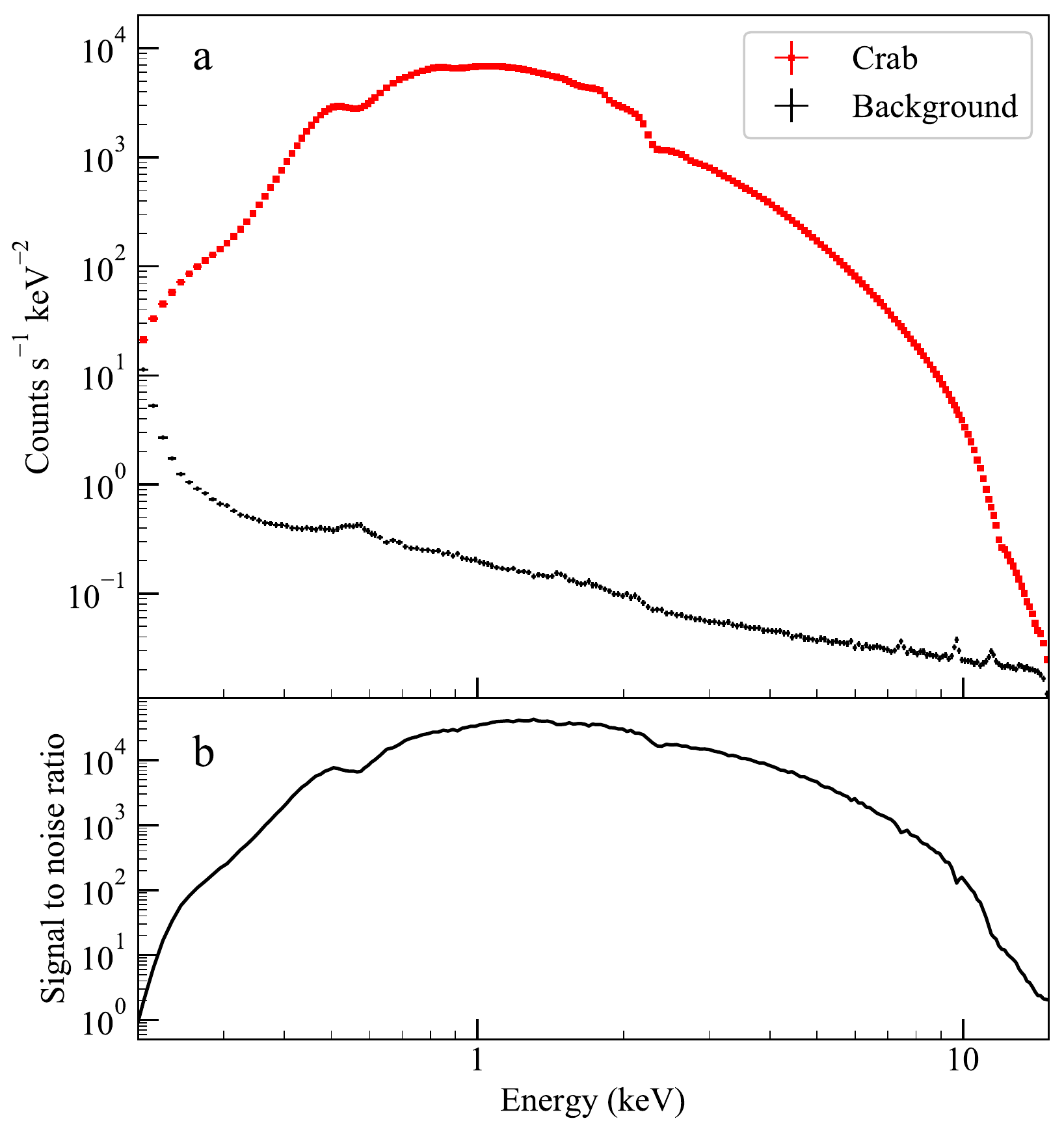}
  \caption{{\bf {\it NICER} X-ray spectrum of the Crab Nebula and Pulsar}. a. The Crab spectrum (red) is compared with a simulated background model from the ``nicer\_bkg\_estimator" (black) \cite{nicerbackground}. {\bf b.} Signal-to-noise ratio of the target spectrum derived from panel a.}
  \label{fig:crabspec}
\end{center}
\end{figure}

\begin{figure}
\begin{center}
  \includegraphics[scale=0.8]{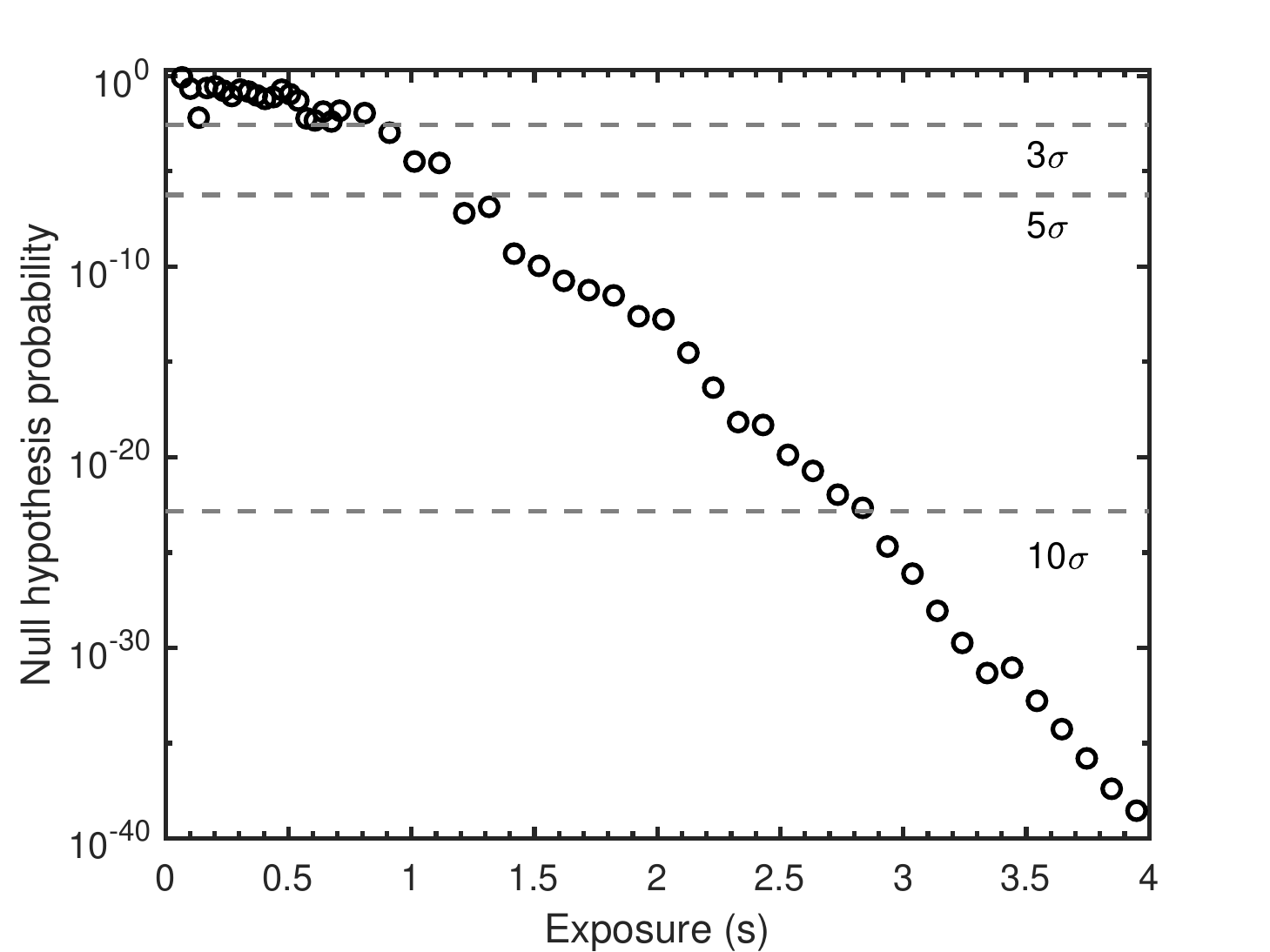}
  \caption{\textbf{Detection significance of the Crab pulsation as a function of the accumulating {\it NICER} exposure}. The  significance is defined as the null hypothesis probability obtained from  chi-square fitting with a constant function to the pulse profile folded at the pulse period listed in Table~\ref{tbl:A2}. The corresponding significance levels (3$\sigma$, 5$\sigma$, and 10$\sigma$) are shown with dashed lines.} 
  \label{fig:crab_pulse_dection_significance}
\end{center}
\end{figure}

\begin{figure}
\begin{center}
  \includegraphics[scale=0.7]{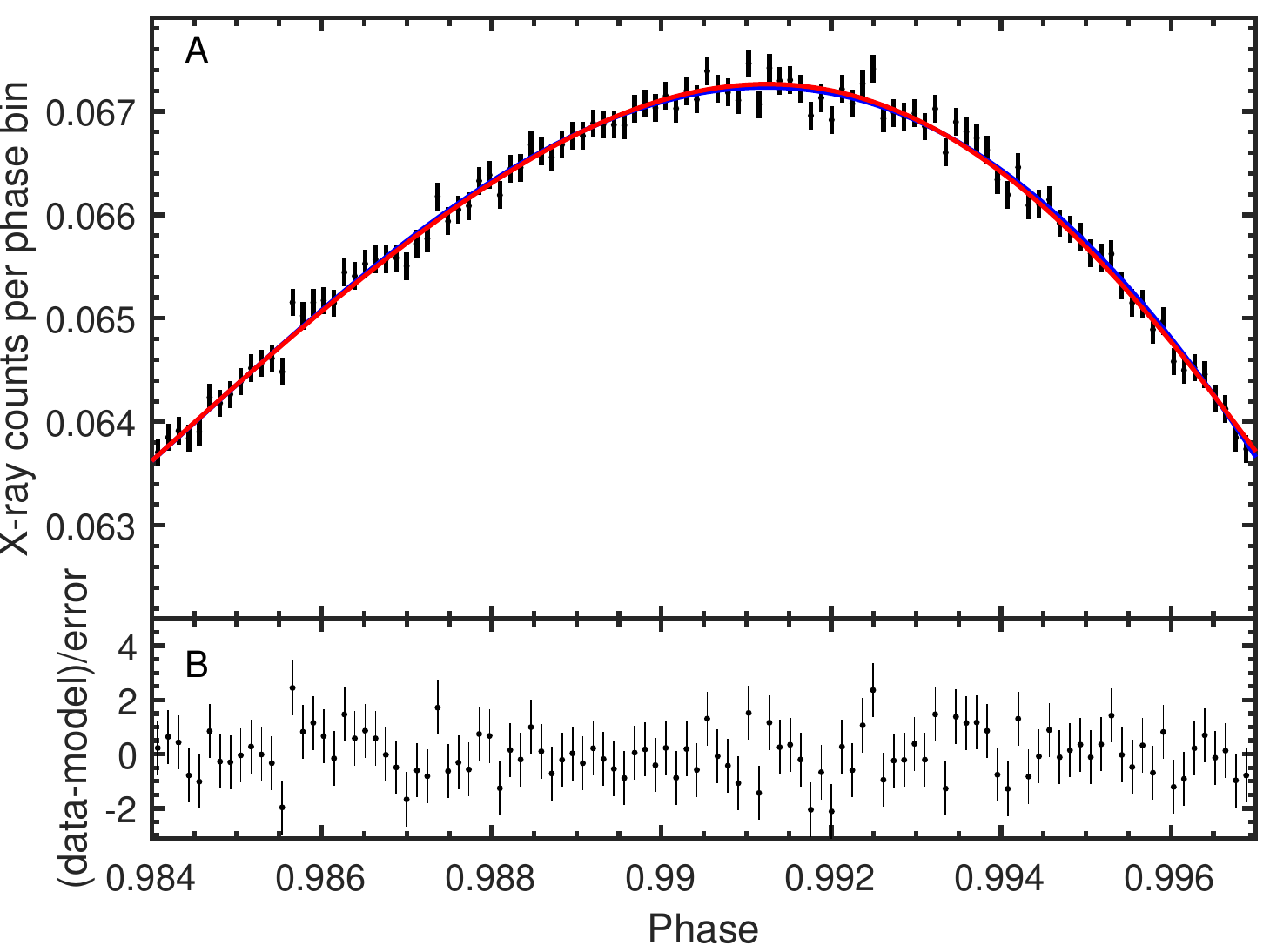}
  \caption{\textbf{MP X-ray pulse profile of the Crab Pulsar observed with {\it NICER}}. A: The data from all the observations are stacked into 8192 phase bins (black points) and fitted with a Fourier series (red line) with 100 harmonics and a 4-th order polynomial function (blue line). The error bars indicate the 1~$\sigma$ statistical uncertainties. B. The residuals with respect to the best-fitting Fourier model. The error bars show 1$\sigma$ statistical uncertainties. } 
  \label{fig:profile_polyfit}
\end{center}
\end{figure}

\begin{figure}
\begin{center}
  \includegraphics[scale=0.8]{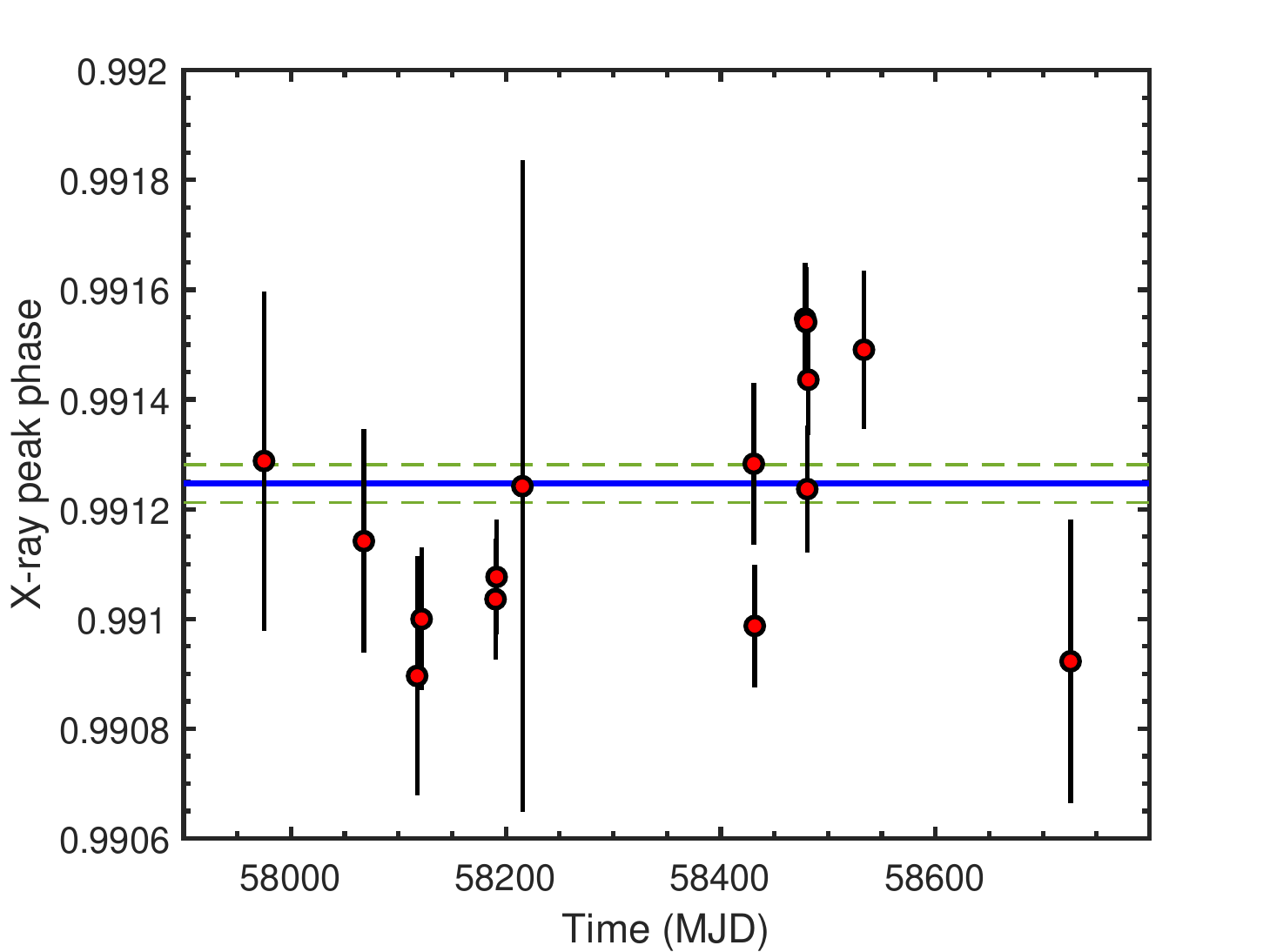}
  \caption{\textbf{MP peak phase as a function of the observation date.} Each pulse profile is fitted using 4096 bins, in the same way as in Figure~\ref{fig:profile_polyfit}. Red dots are best-fitted X-ray peak phase values and with 1$\sigma$ uncertainties. Blue solid and green dashed lines  indicate the X-ray peak  determined from all the stacked profile and its 1$\sigma$ interval, respectively. } 
  \label{fig:xray_center}
\end{center}
\end{figure}

\begin{figure}
\begin{center}
 \includegraphics[scale=0.66]{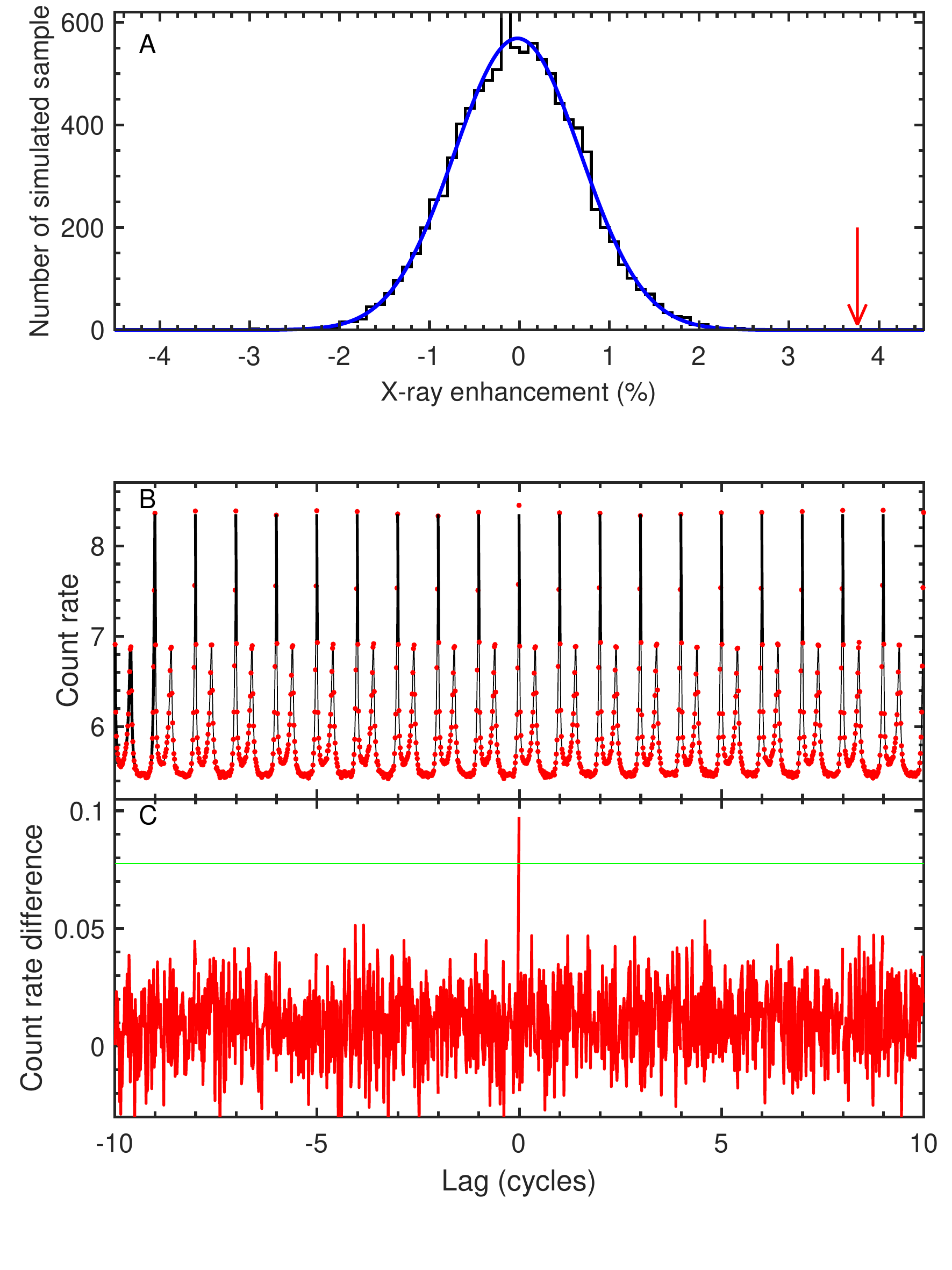}
  \caption{ \textbf{The detection significance of the X-ray enhancement and the lag analysis.}
{\bf A:} Comparison of the $10^4$-sample simulated histogram (in black) of the X-ray enhancements and its best-fitting Gaussian model profile (in blue; mean$=-0.02$\%, $\sigma=0.70$\%) with the measured X-ray enhancement  ratio (indicated with a red arrow; 3.8\%) of the observed data. The X-ray enhancement has a 5.4$\sigma$ significance.
{\bf B:} A series of the (red) Crab pulse profile,  in X-ray counts per phase bin, around the MP-GRPs obtained with the ``lag-analyses" over $\pm$10 pulse cycles, where 64 phase bins are used per rotation, overlaid with the (black) regular pulse profile. Lag $=$ 0 corresponds with phase $=$ 1 defined in equation (\ref{eq:phase_determination}).
{\bf C:} Difference between the data around each MP-GRP-associated profile (red in panel b) and  regular pulses (black in panel b). The green horizontal line denotes the 5$\sigma$ intervals of the differences. X-ray enhancement occurs only in the pulse with the GRP. 
} 
  \label{fig:simulated_histogram}
\end{center}  
\end{figure}

\begin{figure}
\begin{center}
\includegraphics[width=0.99\textwidth]{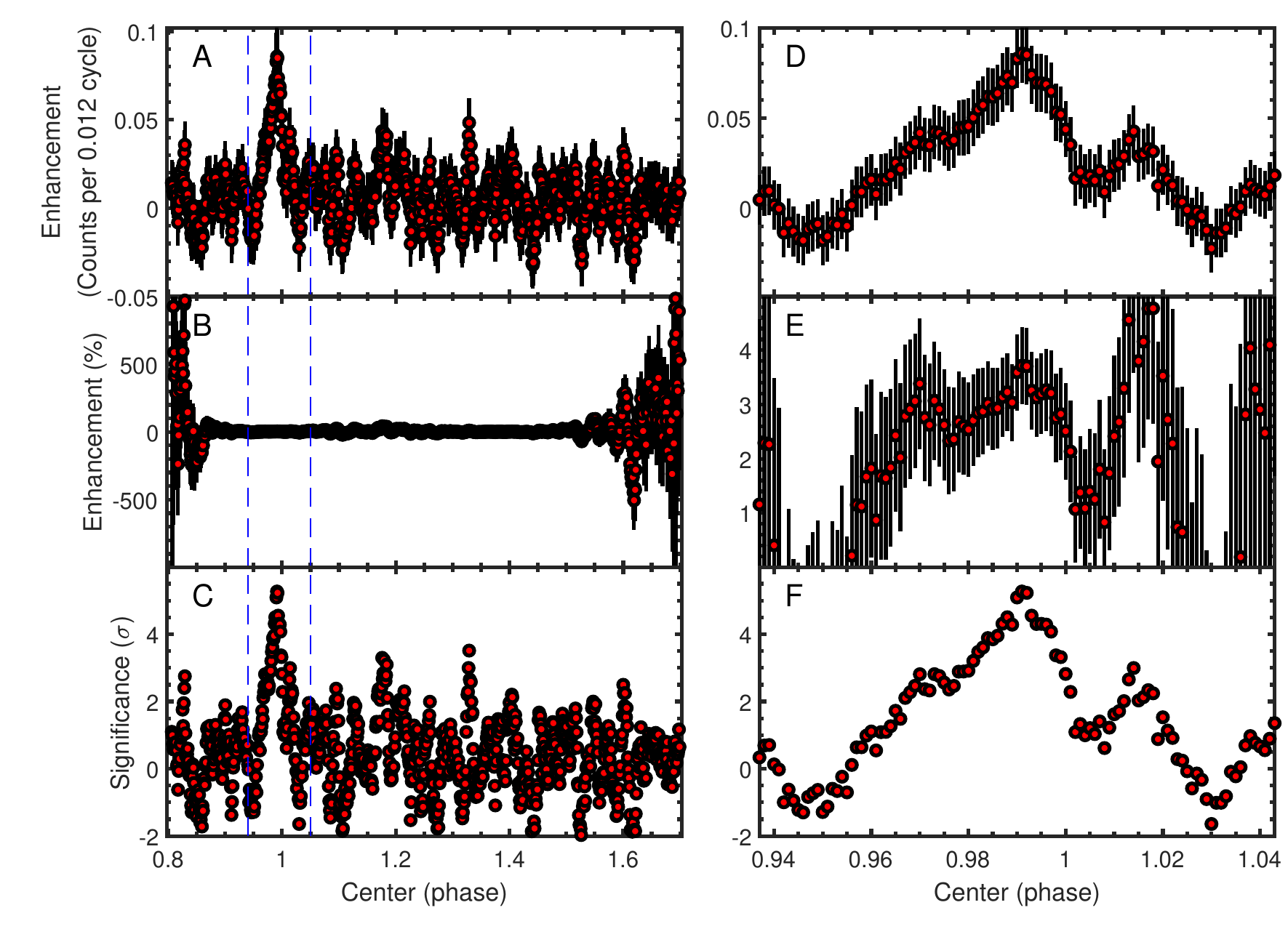}
  \caption{\textbf{The enhancement versus the center of the trial MP phase interval.} Panel A to C show X-ray count rate enhancement, the relative enhancement, and corresponding significance as a function of the trial phase center with a fixed width of 0.012 and spans over the entire pulse cycle except for the off-pulse phases. Panel D to F shows the phase range near the X-ray peak within blue dashed lines in panel A--C. The error bars indicate the 1~$\sigma$ statistical uncertainties. } 
  \label{fig:width_vs_enhancement}
\end{center}
\end{figure}

\begin{figure}
\begin{center}
 \includegraphics[scale=0.66]{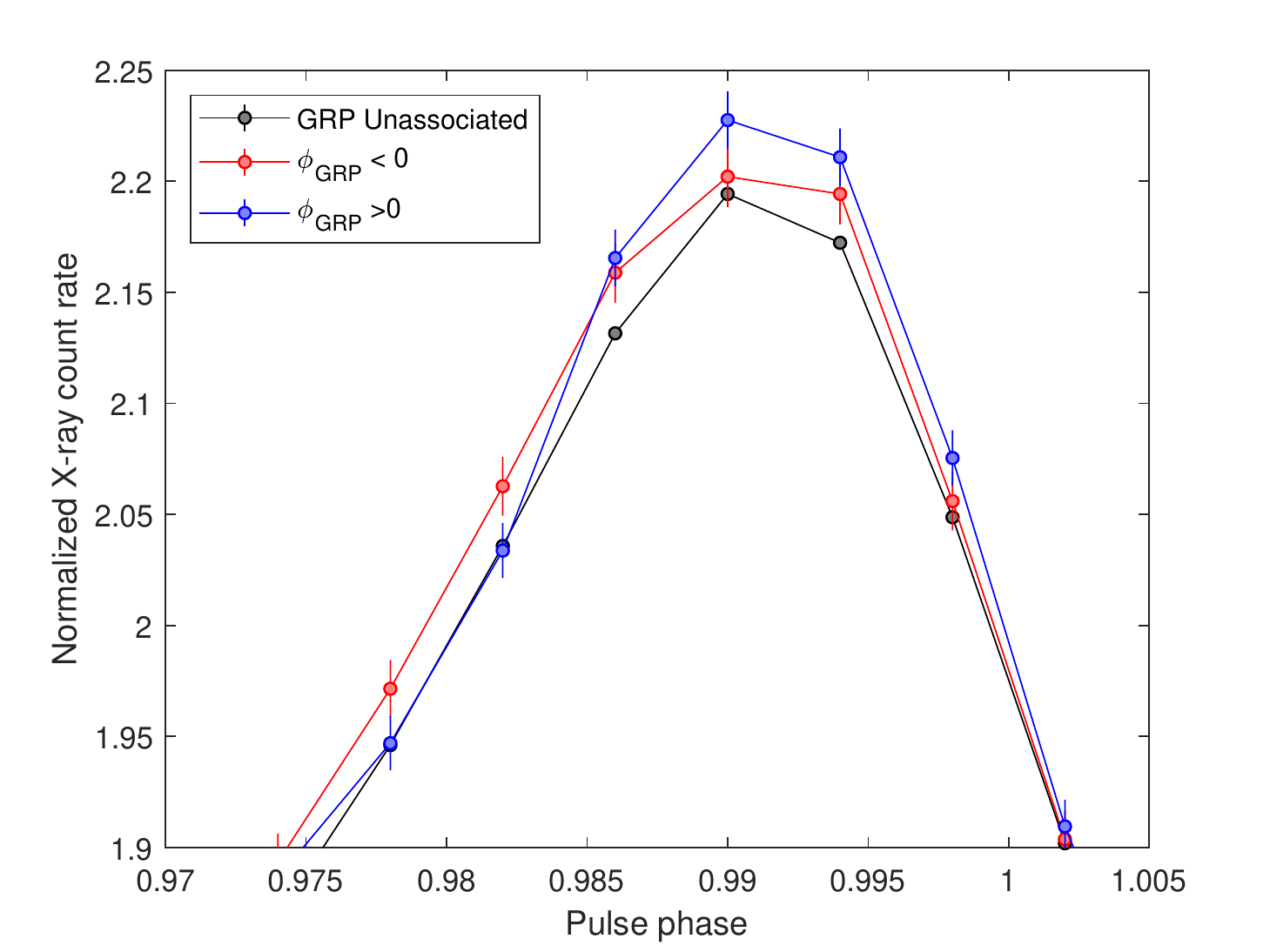}
  \caption{
\textbf{X-ray pulse profiles of pulse cycles unassociated with GRPs (black), cycles associate with $\phi_{\rm{GRP}}<0$ (red), and cycles with $\phi_{\rm{GRP}}>0$ (blue) near the X-ray MP with 1~$\sigma$ statistical uncertainties.} The X-ray count rate is normalized in the same way as Figure \ref{fig:pulse_profile}. 
} 
  \label{fig:profile_diff_grp_phase}
\end{center}  
\end{figure}

\begin{figure}
\begin{center}
 \includegraphics[scale=0.66]{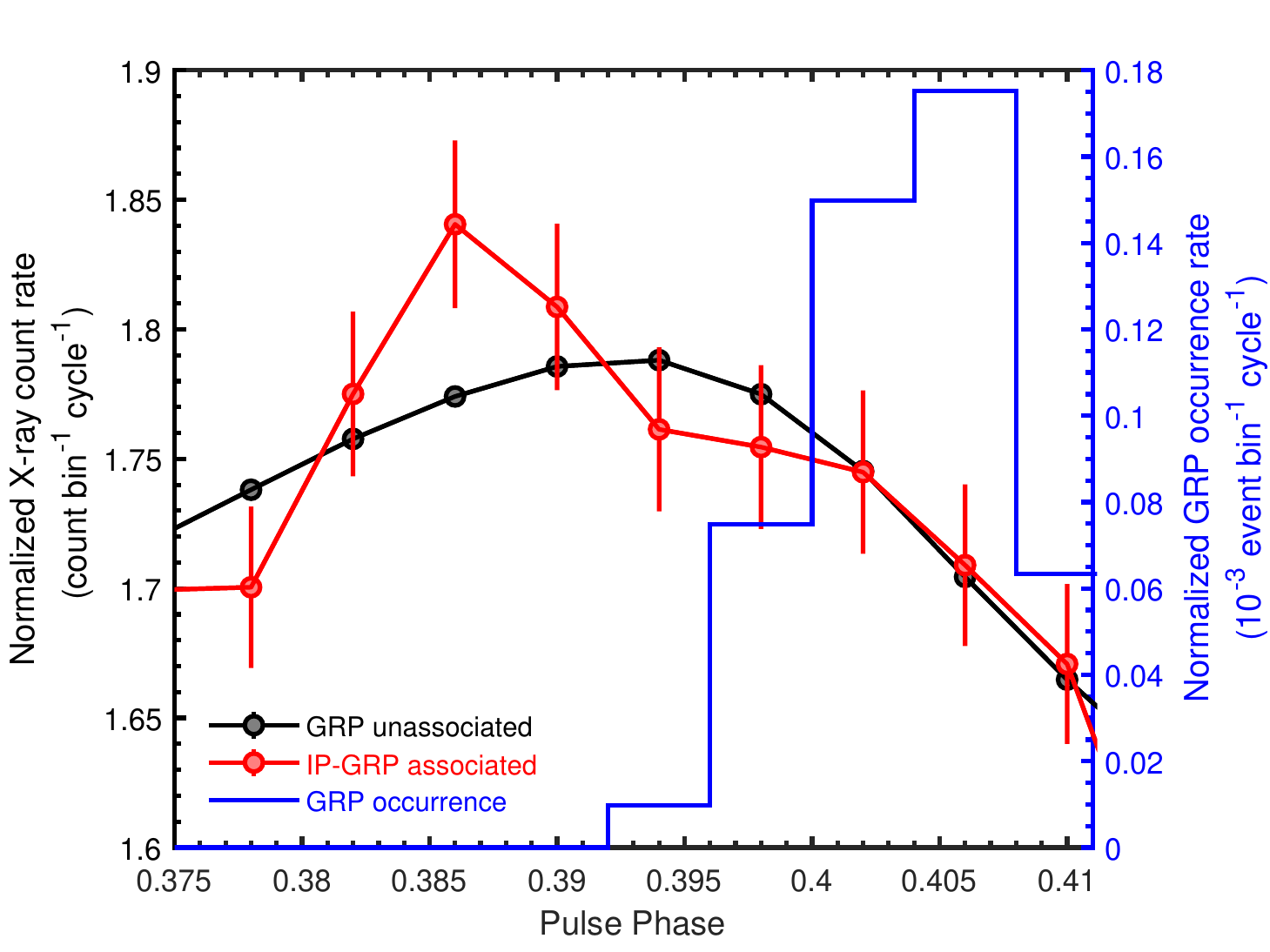}
  \caption{
\textbf{Pulse profile near the IP.} Black and red points connected with solid lines show the X-ray profiles without and with IP-GRP association, respectively. The error bars indicate the 1~$\sigma$ statistical uncertainties. The blue histogram shows the IP-GRP-occurrence distribution.
} 
  \label{fig:profile_ip_grp}
\end{center}  
\end{figure}

\begin{table}[h!]
\begin{center}
\vspace{1mm}
\caption{\textbf{Results of the spectral analysis of MP-GRP-associated and non-GRP-associated spectra.}}
\label{tab:spec_results}
\begin{tabular}{lrr}
\hline \hline
Parameter & MP-GRP-associated & Non-GRP-associated \\
\hline
$N_H$ ($10^{22}$~cm$^{-2}$) &  \multicolumn{2}{c}{0.326$\pm$0.003}  \\ 
O abundance (Solar) &  \multicolumn{2}{c}{0.657$\pm$0.010} \\ 
Fe abundance (Solar) &  \multicolumn{2}{c}{0.458$\pm$0.050} \\
$\alpha$  &1.564$\pm$0.029 & 1.528$\pm$0.016  \\ 
$\beta$   & 0.189$\pm$0.055 & 0.241$\pm$0.014 \\
$K$ (photons~keV$^{-1}$~cm$^{-2}$~s$^{-1}$ at 1\,keV)       & 3.93$\pm$0.06 & 3.81$\pm$0.04  \\ 
0.2-12 keV flux (erg~s$^{-1}$~cm$^{-2}$) & $(2.25\pm 0.04)\times 10^{-8}$& $(2.171\pm 0.004)\times 10^{-8} $ \\ 
\hline
$\chi^2_{\nu}$ (degree of freedom) & \multicolumn{2}{c}{1441.64 (1477)} \\ 
Null hypothesis probability for the joint fitting & \multicolumn{2}{c}{0.740} \\ 
\hline
\end{tabular}
\end{center}
\end{table}

\begin{figure}
\begin{center}
  \includegraphics[scale=0.4]{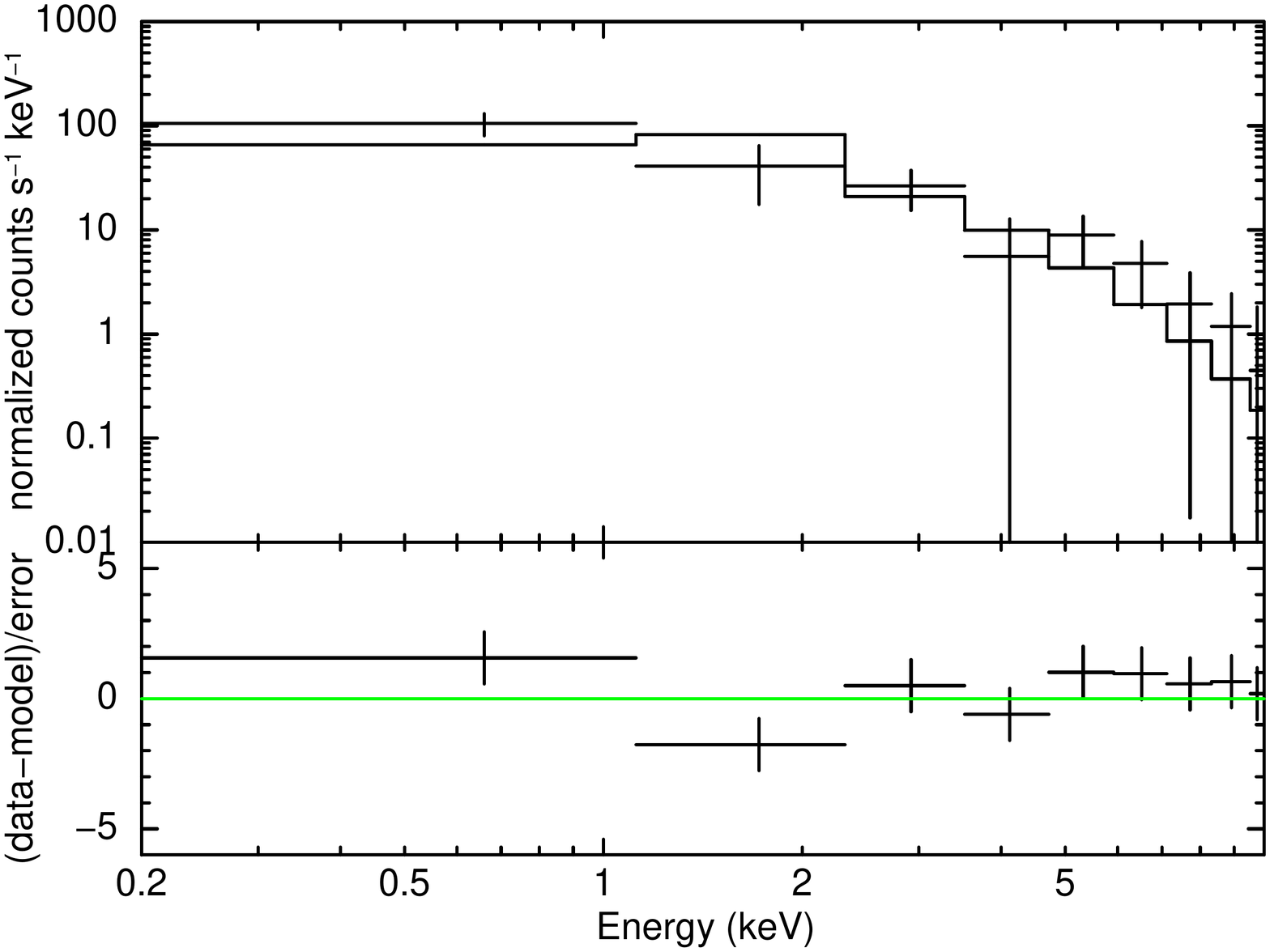}
  \caption{\textbf{The net spectrum of the MP-GRP-associated enhancement.} \textbf{Top panel:} The net MP-GRP-associated enhanced spectrum (black points) using the Non-GRP-associated data as a background. The spectrum is fitted using an absorbed power-law model (histogram). {\bf Bottom panel:} The residual of the top panel between the data and model. See the text for details.}
  \label{fig:netfit_spec}
\end{center}
\end{figure}

\begin{figure}
\begin{center}
  \includegraphics[scale=2.0]{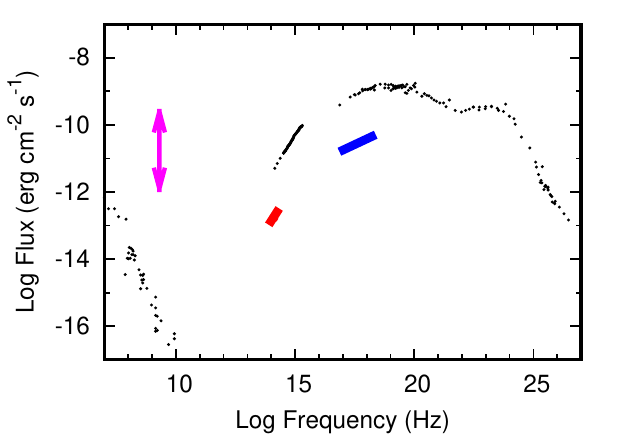}
  \caption{\textbf{Spectral energy distribution (SED) of the persistent emission of the Crab Pulsar  \cite{2014RPPh...77f6901B,2016A&A...591A.134B,2016A&A...585A.133A} of the enhancement during an MP-GRP.} The black dots denote the SED of the persistent emission, where the magenta arrow, red line, and bue line are SED of the enhancement during an MP-GRP in the radio (this work), optical (\cite{2003Sci...301..493S,2013ApJ...779L..12S}), and X-ray (this work) bands, respectively. The GRP peak flux was observed to vary from $10^2$~Jy to $3\times 10^4$~Jy (Figure~\ref{figure:A6}), unlike those in the optical or X-ray bands; the vertical elongation of the magenta arrows represents the range of the radio variation.}
  \label{fig:SED_suppl}
\end{center}
\end{figure}

\begin{figure}
\begin{center}
  \includegraphics[scale=0.7]{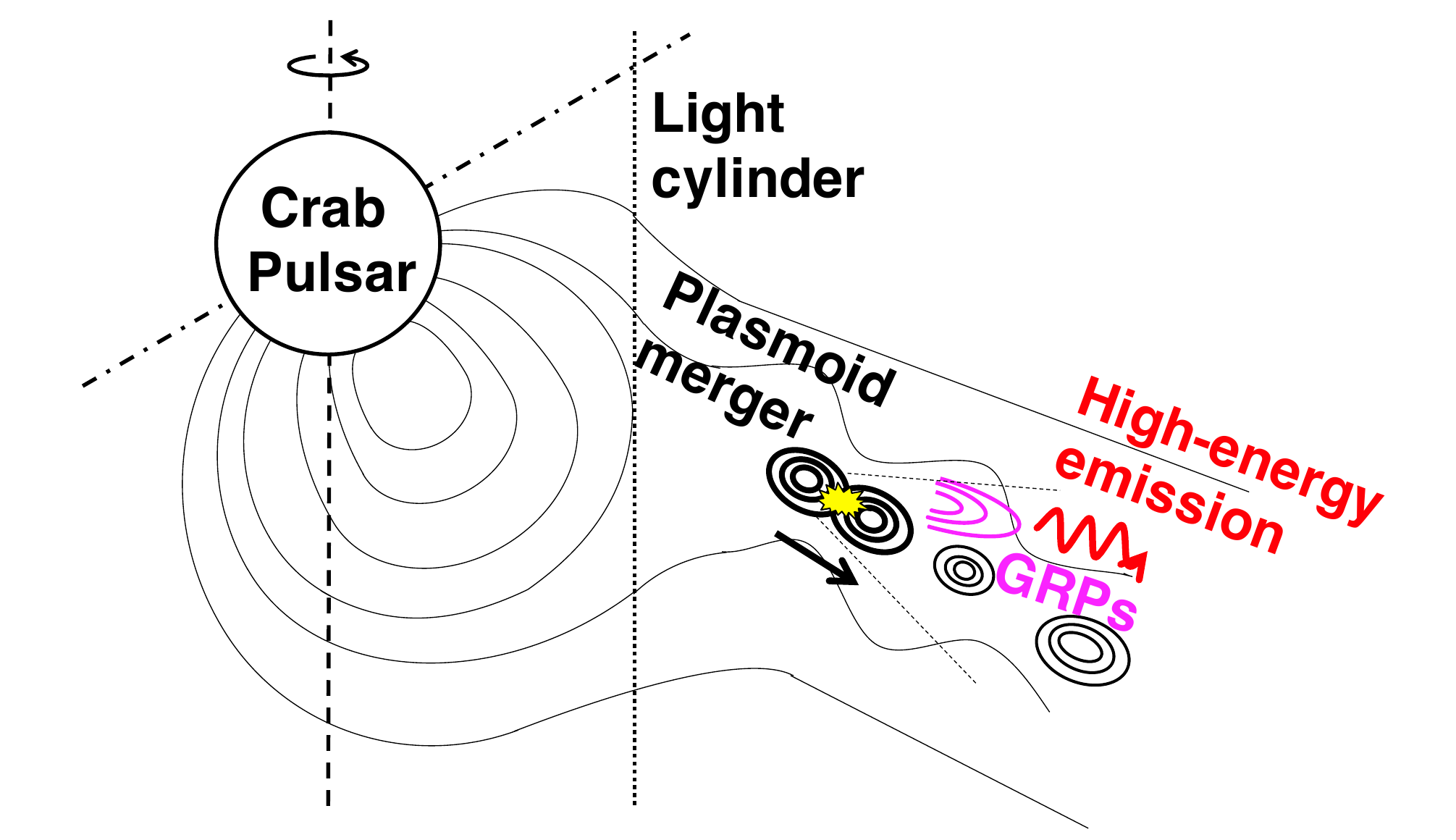}
  \caption{\textbf{Schematic view of the magnetic reconnection model.} The thin curves show the magnetic field lines. Beyond the light cylinder (thick dotted line) the current sheet is fragmented into multiple plasmoids (small loops), which then  undergo successive mergers (yellow point). Plasmas in  merged plasmoids (thick loops) are heated and emit high-energy emission (optical to X-rays) via synchrotron radiation (red). Plasmoids also emit fast magnetosonic waves, which could be observed as GRPs (magenta). The emission is relativistically beamed due to the bulk motion of the merged plasmoids. Thick arrow is the moving direction and the thin dotted line is the beaming angle. Thick dashed and dash-dotted lines show the rotation and dipole magnetic axes of the Crab Pulsar, respectively (thin arrow is the rotation direction).}  
  \label{fig:interpretation}
\end{center}  
\end{figure}

\newpage
\subsection*{Caption for Movie S1}
Movie of 0.3--10.0~keV X-ray pulse profile of the Crab Pulsar observed with {\it NICER} when accumulating exposure time. The profile is generated with 250 phase bins per spin period, includes the contribution from the Crab Nebula, and is normalized by the total number of pulsar spin cycles. Two pulse cycles are shown for clarity. The error bars indicate the 1~$\sigma$ statistical uncertainties. The accumulated number of the pulsar rotation cycles, X-ray events, and exposure, are shown in the title. The detection significance of the pulsation in this plot corresponds with Figure~\ref{fig:crab_pulse_dection_significance}.

\end{document}

%% file: suppl_01_previous_studies_v2.tex
\subsection*{Previous searches for high-energy emission associated with GRPs}
\label{Previous high-energy studies associated with GRPs}

In this section, we review prior studies of GRPs in X-rays and gamma-rays from the Crab Pulsar. 
The upper limits derived from these observations are listed in Table \ref{tab:table_previous_observations}.

\subsubsection*{X-rays}
The search for a variation of the pulsed emission has been reported in several studies \cite{1993AIPC..280..213K}. 
One study \cite{2012ApJ...749...24B} performed 
5.4\,hours of simultaneous observations of the Crab Pulsar with the Green Bank Telescope (GBT) at 1.5\,GHz and the Chandra X-ray satellite at energies from 1.5 to 4.5\,keV. They searched 
for an X-ray flux enhancement a) in the rotations during which GRPs occurred [for main pulse (MP) and interpulse (IP) GRPs separately], b) in the time windows around GRPs ranging from one up to 15 pulsar periods and c) during the GRPs that occurred simultaneously in both the X-ray and radio bands. In all three cases, no significant correlations  were found, the results being compatible with 2\,$\sigma$ fluctuations.

A correlation search at hard X-rays from the Crab Pulsar was carried out with the Suzaku X-ray observatory simultaneously with the Kashima and Usuda radio telescopes \cite{2015PhDT.......M}. They  found no significant flux increase in the 15--75 keV band (silicon PIN diodes of the Hard X-ray Detector; HXD-PIN) coincident with any of the GRPs and derived 95\%-confidence upper limits of potential enhancements of 33\% and 88\% for the MP-GRPs and  IP-GRPs, respectively. Similarly, they  derived 95\%-confidence upper limits of 63\% and 193\%  in the 35--315~keV band (gadolinium silicate scintillators [Gd$_2$SiO$_5$(Ce)) of the HXD; HXD-GSO. 

Correlation studies were carried out with the Hitomi X-ray satellite in an energy range from 2 to 300\,keV and the Kashima radio telescope at 1.4 to 1.7\,GHz \cite{2018PASJ...70...15H}.  On the basis of the detection of 3,090 MP- and 260 IP-GRPs (1,000 MP- and 100 IP-GRPs were detected simultaneously)  from about 2\,ks of observations, 
searches for a) X-ray profile changes before and after GRP events and b) X-ray peak enhancements during GRP events were carried out. All examined variations were reported to be within  2\,$\sigma$ fluctuations of the X-ray fluxes at the pulse peaks.

\subsubsection*{Gamma-rays}
A search for a correlation of GRPs with gamma-ray photons was carried out by \cite{1974NCimB..24..153A}. They observed the Crab pulsar simultaneously with the radio telescope of the Dominion Radio Astrophysical Observatory at 146 MHz and with the Whipple Imaging Air Cherenkov Telescope (IACT) at energies above 200\,GeV for about 10\,hours, and detected 300 GRPs.  The time resolution of the observations was 200\,\textmu s, which enabled them to carry out correlation searches at time scales of one to five rotation periods of the Crab pulsar. No statistically significant enhancement was found.

 Correlation searches of GRPs with gamma-ray photons  have been carried out predominantly with spacecraft. Simultaneous observations with the 43\,m Green Bank radio telescope at 0.8 and 1.4\,GHz and the Oriented Scintillation Spectrometer Experiment (OSSE) on the Compton Gamma-Ray Observatory (CGRO) in an energy range from 50 to 220\,keV  resulted in a collection of about 3,600 GRPs. 
  No increases of the gamma-ray flux  by more than a factor of 2.5 were detected \cite{1995ApJ...453..433L}. 

Further correlation searches were carried out with the GBT at 0.80, 0.81, and 1.33\,GHz and with the Energetic Gamma Ray Experiment Telescope (EGRET) onboard  the CGRO at energies higher than 50\,MeV \cite{1998ApJ...496..863R}. 
On timescales ranging from 6 to 30 pulsar rotations, no enhancement of the gamma-ray flux at energies above 50\,MeV was found.

Searches for correlations between GRPs and gamma-ray photons were also carried out with 10\,h-long simultaneous observations of Fermi Large Area Telescope (Fermi-LAT) and GBT \cite{2011ApJ...728..110B,Mickaliger_2012}, targeting  a specific class of GRPs that occur at the phase ranges of the High Frequency Interpulse Component, HFIP \cite{1996ApJ...468..779M,2016ApJ...833...47H}, as predicted by a theoretical model 
\cite{2007MNRAS.381.1190L}.   
They observed $2.1\times 10^{4}$ GRPs at 8.9 GHz and 77 gamma-ray photons with energies between 100\,MeV and 5\,GeV and searched for a correlation between the  occurrence frequency of GRPs and single gamma-ray photons, as well as changes of gamma-ray flux  during and around the period of  each GRP.
No significant changes in the GRP generation rate were  detected in time windows ranging from 10 to 120 seconds, with a 95\% upper limit for a gamma-ray flux enhancement in a pulsed phase emission window around the observed GRPs of
four times the average pulsed gamma-ray flux of the pulsar.

Studies using Fermi-LAT \cite{Mickaliger_2012} used 107\,h of simultaneous observations at 0.33 and 1.2 \,GHz with the 43\,m radio telescope in Green Bank, where  the energy of the gamma-ray data ranged between 0.1 and 100\,GeV.  They examined correlations and anti-correlations of 92,022 detected (MP and IP) GRPs with regard to their 393 gamma-ray photons (at time lags of $\pm 3\times 10^{6}$ rotation periods), collected  from the period  before and after the Crab Nebula flare, a sudden increase of unpulsed
gamma-ray flux from the nebula observed in October
2007 and September 2010 \cite{2011Sci...331..736T}. They determined the largest deviations from the mean of random correlations to be 2.4\,$\sigma$ and 2.1\,$\sigma$ for  anti-correlated post-flare 1.2\,GHz MP GRPs and correlated pre-flare 1.2\,GHz IP GRPs, respectively, at a time lag of 20,000 rotation periods. 
They concluded that these were not statistically significant.

Correlation searches were carried out with the Very Energetic Radiation Imaging Telescope Array System (VERITAS) at very high energy (VHE) gamma-rays above 150\,GeV and the GBT at a frequency of 8.9\,GHz \cite{2012ApJ...760..136A} and investigated HFIP GRPs. 
Using 15,366 GRPs and 30,093 gamma-ray photons, they searched the detected MP and IP GRPs for an enhancement of pulsed gamma-ray emission in eight-time windows with  durations from one up to 2,182 rotation periods at three locations with regard to a GRP to account for leading, lagged and concurrent correlations. They reported upper limits of the gamma-ray flux of the Crab Pulsar during HFIP GRPs that were detected simultaneously with the radio data to be five to ten times the average  flux and two to three times the average gamma-ray flux on time scales of about eight seconds around HFIP GRPs.

A study covering energies above 60\,GeV, was carried out 
using the Major Atmospheric Gamma-ray Imaging Cherenkov telescopes (MAGIC) and the Effelsberg radio telescope and the Westerbork Synthesis Radio Telescope (WSRT) at 1.3\,GHz \cite{2020A&A...634A..25M}.  
They observed 99,444 GRPs and 433 gamma-ray photons from 16\,hours of overlapping observations, 
and performed correlation searches of MP and IP GRPs at time windows with durations ranging from 1/9 to 2,187 rotation periods and three different locations with respect to  each GRP,  following the approach of \cite{2012ApJ...760..136A}. 
The authors reported no  statistically significant correlation between GRPs and gamma-ray photons for time windows of 1/9, 1/3, 1 and 3 rotation periods, 
reporting upper limits of 12\% to 2,900\%  on the flux increase of the pulsed gamma-ray flux during a radio GRP.

%% file: suppl_02_radio_ver0204_v2.tex
\subsection*{Radio data and analysis}
\label{suppl:radio}
\subsubsection*{Radio observations}\label{suppl:A-1}

Radio observations of the Crab Pulsar were 
performed using radio telescopes at 
Usuda and Kashima (Table \ref{tbl:A1})  in the S and X bands for right hand circular polarization   
with eight channels in total: four S-band channels (ch0--3: 2,194--2,226~MHz, 2,226--2,258~MHz, 2,258--2,290~MHz, and 2,290--2,322~MHz) 
and four X-band channels 
(ch4--7: 8,374--8,406~MHz, 8,406--8,438~MHz, 8,438--8,470~MHz, and 8,470--8,502~MHz) 
and a total data rate of 2.048~Gbits/s (= 8 channels $\times$ 4 bits $\times$ 64M samples/s).
We only report the S-band data because  
there are not enough GRPs in the X band data. 
While we analyze all four S-band channels of the Usuda data, 
we only consider three S-band channels (ch0, ch2, and ch3) of the Kashima data 
as one channel (ch1) was severely affected by 
radio frequency interference (RFI). 
In Table \ref{tbl:A2}, all observing sessions are tabulated with the parameters used in the analysis.

\subsubsection*{De-dispersion and an example of GRP}\label{suppl:A-2}
In this subsection, we present example data obtained using the Usuda radio telescope on MJD 58121 (session 4).
Fig. \ref{figure:A1}A shows the squared raw-antenna voltages in channel 3, 
$|V^{\rm raw}_{\rm ch3} (t)|^2$, 
for a 3 ms duration, along with their 
10~\textmu s averages. 
A dynamic spectrum (Fig. \ref{figure:A1}B)
shows a down-tone chirp signal, which indicates arrival of a pulse
with a frequency dispersion, or a group delay, caused by interstellar free electrons.
This frequency dispersion can be removed by a coherent de-dispersion process \cite{1971ApJ...169..487H, 2004hpa..book.....L},
which is described as,
\begin{equation}
V^{\rm de\mathchar`-dis}_{\rm ch3} (t) = \int df e^{i2\pi (f_0+f) t} H^*(f_0+f) \int dt' e^{-i2\pi (f_0+f) t'} V^{\rm raw}_{\rm ch3}(t'),
\end{equation}
where $V^{\rm de\mathchar`-dis}_{\rm ch3}(t)$ is antenna voltages after de-dispersion and
$H^*(f_0+f)$ is the complex conjugate of a transfer function $H(f_0+f)$ with a base frequency $f_0$ (2,322 MHz for ch3) and an offset $f$,
\begin{equation}
H^*(f_0+f)=\exp \left \{  -i \frac{2\pi {\cal D}~ {\rm DM} f^2} {(f_0+f)f_0^2} \right \},
\end{equation}
where ${\cal D} = \displaystyle{\frac{e^2}{2\pi m_e c}}=4.148808 \times 10^{9}$ MHz pc$^{-1}$ cm$^3$
($e$, $m_e$, and $c$ are electron charge, electron mass, and speed of light, respectively)
and DM$=$56.7548 pc~cm$^{-3}$.
In Fig.~\ref{figure:A2},  squared values $|V^{\rm de\mathchar`-dis}_{\rm ch3} (t)|^2$ for a 0.1 ms  duration
(i.e. zooming into part of Fig.~\ref{figure:A1}A after de-dispersion) indicate a pulse with a few \textmu s duration (this pulse is identified as an MP GRP in our subsequent analysis; see Spin number, phase, and GRP identification subsection). 
With a time constant $\Delta t=10~$\textmu s, we take the average
\begin{equation}
\overline{|V^{\rm de\mathchar`-dis}_{\rm ch3}(t_k)|^2} = \frac{1}{\Delta t} \int^{t_{k+1}}_{t_k} dt~ |V^{\rm de\mathchar`-dis}_{\rm ch3} (t)|^2,
\end{equation}
for binned times $t_k=t_{\rm start} + k \Delta t$~$(k=0,1,2,\cdots)$,
where $t_{\rm start}$ is the start time of the observing session.
The resultant $\overline{|V^{\rm de\mathchar`-dis}_{\rm ch3}(t_k)|^2}$ is  overlaid in Fig. \ref{figure:A2}.

\subsubsection*{Initial data processing}\label{suppl:A-3}
We de-disperse data 
(in each channel $j$)
 from all observing sessions in the same way as  
described in the previous subsection 
and obtain
$\overline{|V^{\rm de\mathchar`-dis}_{\rm j}(t_k)|^2}$ for $t_{k}$ defined for each session. 
Since DM of the Crab Pulsar shows time variations with time scale as short as several days, DM should be evaluated for each session.
We determine DM iteratively so that the sub-microsecond structures of GRPs in  all frequency channels  are aligned to each other 
after the de-dispersion procedure.
For the initial trial value of a DM in each iteration, we use values from the Jodrell Bank ephemeris \cite{Jodrellbank}. 
The estimated uncertainty of DM is $\sim \pm$0.001 $\rm pc~cm^{-3}$,
which corresponds to the uncertainty of the pulse arrival time $\sim \pm $0.76~\textmu s or  
that of the spin phase $\sim \pm 2.3 \times 10^{-5}$.

\par
From $\overline{|V^{\rm de\mathchar`-dis}_{\rm j}(t_k)|^2}$ we make incoherent summations,
\begin{equation}
  {\cal{W}}(t_k) \equiv
 \sum_{j} \overline{|V^{\rm de\mathchar`-dis}_{j}(t_k+\tau_{j3})|^2},
\end{equation}
\noindent
where $j$ runs over the available channels (four channels for Usuda and three channels for Kashima) 
and $\tau_{j3}~(j=0 - 3)$ is a group delay time between channels $j$ and 3 ($\tau_{33}=0$).
For $j\ne 3$, $\overline{|V^{\rm de\mathchar`-dis}_{j}(t_k+\tau_{j3})|^2}$ is linearly interpolated
from the values of $\overline{|V^{\rm de\mathchar`-dis}_{j}(t_{k'})|^2}$  at $t_{k'} > t_k$.

From $ {\cal{W}}(t_k)$ of $\Delta t=10$ {\textmu}s duration, we calculate
1~s averages $\overline{\cal W}(t_N)$ and standard deviations $\sigma (t_N)$,
where $t_N$ $(N=0,1,2,...)$ are integer seconds, i.e.,
\begin{equation}
t_N = t_{\rm start} + N.
\end{equation}

As an example, Fig. \ref{figure:A3} shows  $\overline{\cal W}(t_N)$ and $\sigma (t_N)$ from 9:00:00 UT to 16:20:00 UT of session 4.
In panel A, $\overline{\cal W}(t_N)$ shows a gradual variation ($\sim$2\%), which is caused by the gain drift of the receiving system.
In panel B, there are spikes of $\sigma (t_N)$,
the largest of which (marked a) corresponds to the MP GRP shown in Fig. \ref{figure:A2}.
This short ($\sim 10$ {\textmu}s) GRP boosted the 1~s standard deviation $\sigma(t_N)$ by a factor of $\sim$5.
Another spike (marked b) corresponds to an MP GRP at 11:26:30 UT.
In panel C, which shows $\sigma (t_N)$ with an enlarged vertical scale,
there are numerous spikes, which are all identified as MP/IP GRPs (see Spin number, phase, and GRP identification subsection). 
To remove these spikes for the following signal-to-noise analysis, 
we define the standard deviation not affected by GRPs as $\overline{\sigma} (t_N)$,
which is approximated as $0.026 \overline{\cal W}(t_N)$ (a black line in Panel C).
The signal to noise ratio, SNR$(t_k)$ at $t_k \in [t_N, t_{N+1})$, is then defined as,
\begin{equation}
{\rm SNR}(t_k) = \frac{ {\cal W}(t_k) - {\overline {\cal W}}(t_N) }
                      {                  \overline {\sigma} (t_N) }.
\label{eqn:SNforDeDdata}
\end{equation}

Two GRPs corresponding to the spikes a and b in Fig. \ref{figure:A3}
are found to have peak signal-to-noise ratios of 1712.7 and 799.9, respectively.

\subsubsection*{Time-domain RFI rejection}\label{suppl:A-4}
To eliminate RFI from both Usuda and Kashima data in the time domain,
we calculate summations of the squared raw antenna voltages (without de-dispersion) 
${\cal W}^{\rm raw}(t_k)$ directly from $V^{\rm raw}_j(t)$, i.e.,
\begin{equation}
{\cal W}^{\rm raw}(t_k) \equiv \frac{1}{\Delta t} \sum_j \int_{t_k}^{t_{k+1}} dt ~|V^{\rm raw}_{j}(t)|^2,    
\end{equation}
where $j$ runs over the channels used.                 
We calculate its average  [$\overline {\cal W}^{\rm raw}(t_N)$] 
and standard deviation [$\sigma^{\rm raw}(t_N)$] for each 1~s interval.
If $\sigma^{\rm raw}(t_N)$ exceeds $\sigma(t_N)$, we reject the data ${\rm SNR}(t_k)$ for 
$t_k \in [t_N, t_{N+1})$ and
regard it as being affected by RFI.
After this procedure, we visually inspect all  remaining data and further reject intervals where RFIs appear to still remain. 
We find 99.3\% and 94.0\% of Usuda and Kashima data, respectively, are free from RFI.

\subsubsection*{Spin number, phase, and GRP identification}\label{suppl:A-5}
We use the pulsar timing package {\small TEMPO2} \cite{2006MNRAS.369..655H} to convert $t_k$ to the Solar System barycenter time (TDB) ${\tilde t}_k$.
 From ${\tilde t}_k$, we calculate turn $y_k$, daily sequential spin number $N_{\rm SEQ}$, and spin phase $\phi_k$ according to
\begin{equation}
\begin{array}{ccl}
y_k         & = &y_0 + \nu_0 ({\tilde t}_k-{\tilde T}_0) + 0.5 {\dot \nu}_0 ({\tilde t}_k - {\tilde T}_0)^2 \\ \\
N_{\rm SEQ} & = & {\rm floor} (y_k+0.5)  \\ \\
\phi_k      &=  & y_k - N_{\rm SEQ} +1,
\label{eq:phase_determination}
\end{array}
\end{equation}
where $y_0$ is the initial phase, $\nu_0$ and $\dot{\nu}_0$ are the spin frequency and its time derivative, respectively, all of which
are  determined at ${\tilde T}_0$=00:00:00 TDB of the day of the observing session
 with the interpolated values of the Jodrell Bank monthly ephemeris \cite{Jodrellbank}.
The values of $\nu_0$ and $\dot{\nu}_0$ are tabulated in Table \ref{tbl:A2}.
The planetary ephemeris used by the Jodrell Bank observatory (DE200) differs from that used by the NICER data analysis (DE430).
We checked how the parameters $y_0$, $\nu_0$, and $\dot{\nu}_0$ appearing in equation (\ref{eq:phase_determination})
depend on the choices of DE200/DE430 and found that only the dependence of
 $y_0$ should be taken into account and that those of $\nu_0$ and $\dot{\nu}_0$ are negligible for estimating $y_k$. 
The uncertainty of $y_0$ is at most $7\times 10^{-5}$ (Table \ref{tbl:A2}), 
which is $\sim$2\% of the accumulation bin size for GRP and X-ray photons 
so it has a negligible effect on our analysis.
The mathematical function ${\rm floor} (x)$ for a real number $x$ returns  the integer $N$ satisfying $x-1<N \le x$.
As a result, the range of the phase is guaranteed to be $0.5 \le \phi_k < 1.5$,
where the peak of the average main pulse is  placed at phase $=1$. 

Fig. \ref{figure:A4} shows diagrams of time versus phase for ${\rm SNR}(t_k)$ of  
observing session 4 with four different lower threshold values of the
signal to noise ratio SNR$_{\rm thr}$.
We  find two data clusters at phases of $\sim$1  and $\sim$1.406. 
Data-points from phase ranges $\phi=$ 0.9917--1.0083 and $\phi=$ 1.3944--1.4111 are categorized 
as the MP and IP GRP candidates, respectively [figure 4 of \cite{2004ApJ...612..375C}]. 
If multiple MP/IP GRP candidates at the same $N_{\rm SEQ}$ are found,
we choose the one with the maximum SNR$(t_k)$ as representing the MP/IP GRP at this $N_{\rm SEQ}$.
Outside the MP and IP phase ranges, there are background data-points,
whose distribution does not depend on $\phi$ and 
varies from dense (panel A) to sparse (panel D) depending on SNR$_{\rm thr}$.
By assuming homogeneity over $\phi=0.5-1.5$, 
we calculate the pseudo event density per unit $\phi$, $\rho_{\rm bg} (\rm{SNR}_{\rm thr})$,
and obtain the expected value of the false GRP number, $N_{\rm false}=\rho_{\rm bg} (\rm{SNR}_{\rm thr}) \Delta \phi$,
where $\Delta \phi$ is a phase width (0.0167 for both MP GRP and IP GRP).
If we observe $N^{\rm MP}_{\rm obs}$ candidates for MP GRP with SNR$(t_k)>$SNR$_{\rm thr}$,
the expected value of a true MP GRP number 
is $N^{\rm MP}_{\rm obs}-N_{\rm false}$. 
Let $N^{\rm MP}_{{\rm obs},i}$ and $N_{{\rm false},i}$  be
the numbers defined above for session $i$.
False-positive MP GRP rates of Usuda and Kashima data are calculated as
\begin{equation}
\left .
\frac{\displaystyle{\sum_{i\in {\rm [U]}}} N_{{\rm false}, i} } 
     {\displaystyle{\sum_{i\in {\rm [U]}}}          N^{\rm MP}_{{\rm obs},i}}
\right |_{\rm{SNR}_{\rm thr}}
~~~{\rm and}~~~
\left .
\frac{\displaystyle{\sum_{i\in {\rm [K]}}}  N_{{\rm false},i}   }
     {\displaystyle{\sum_{i\in {\rm [K]}}}          N^{\rm MP}_{{\rm obs},i}}
\right |_{\rm{SNR}_{\rm thr}} ,
\end{equation}
respectively, where 
$[\rm U]=[1,2,3,4,8,9,13,15]$ 
and 
$[\rm K]=[5,6,7,10,11,12,14]$.
Similarly, false-positive IP GRP rates of Usuda and Kashima data are calculated.
Table \ref{tbl:A3} lists the numbers of GRP candidate and false-positive GRP rates,
both of which are decreasing functions of SNR$_{\rm thr}$. 
Because the sensitivity  of the Kashima observatory is $\sim$55\% that of the Usuda observatory, 
the false GRP rates for Kashima are higher than those for Usuda. 
There are two competing factors;
it is desirable 
(1) to have as much observed numbers of GRP candidates and
(2) to reduce false GRP rates. 
Balancing these two factors, we choose SNR$_{\rm thr}$=5 for the following analysis.

If a GRP occurs at around the time between ${t}_k$ and ${t}_{k+1}$, 
its contribution to the squared voltage is  split into ${\cal W}(t_k)$ and ${\cal W}(t_{k+1})$, and the peak intensity is artificially lowered.
To compensate for this effect, we further perform a half-shifted-bin procedure  
(see section A.1.4 of \cite{2018PASJ...70...15H}).
In the worst case, SNR$(t_k)$ at the peak of a GRP becomes lower than SNR$_{\rm thr}$, and accordingly this GRP is miscounted.
 This type of miscount is handled with this procedure.

\subsubsection*{GRP and regular pulse shapes}\label{suppl:A-6}
The phase ranges of GRPs of the Crab Pulsar
coincide with those of the regular pulses
[e.g., \cite{2008ApJ...676.1200B, 2004ApJ...612..375C}].
We confirmed this property
by comparing the number of GRPs and regular pulse shapes for all observing sessions.
An example (from  observing session 4) is shown in Figs. \ref{figure:A5}A and B for MP GRPs and in C and D for IP GRPs.
 We find that the occurrence of GRPs is limited to the phase ranges
of the corresponding regular pulses.

\subsubsection*{Temporal variations of the GRP detection rate}\label{suppl:A-7}

Table \ref{tbl:A4}  summarizes our GRP identification for all observing sessions.
The temporal variations of the GRP detection rates are usually interpreted as an effect of  interstellar scintillation \cite{1995ApJ...453..433L}. 
These observing sessions include the epoch of a large glitch in the Crab Pulsar rotation rate \cite{2018MNRAS.478.3832S}. 
We checked whether this glitch affected the GRP rate. 
In Fig. \ref{figure:A6},
we compare the variations of the GRP rate with  variations of the spin rate change ($\dot{\nu}_0$) 
taken from the Jodrell Bank ephemeris \cite{Jodrellbank}.
We find no change in the GRP rates with respect to $\dot{\nu}_0$.

\subsubsection*{Flux density and fluence}\label{suppl:A-8}
Because observations did not involve the use of switching noise diodes for absolute flux calibration, 
we estimate flux density $I({ t}_k)$  
using the radiometer equation \cite{1946RScI...17..268D, 2004hpa..book.....L}
\begin{equation}
I({{t}}_k) = {\rm SNR}({t}_k) \times \frac{\mathrm{SEFD}+S_{\rm CN}} {\sqrt{\Delta \nu \Delta t~n_{\rm p}}}~~~~{\rm [Jy]},
\label{eqn:radiometer}
\end{equation}
where
SEFD (system equivalent flux density) and $\Delta \nu$ (effective band width) are tabulated in Table \ref{tbl:A1}.
$\Delta t$ is the integration time (10~\textmu s), and 
$n_{\rm p}$ is the number of polarizations observed ($n_{\rm p}=$1). 
$S_{{\rm CN}}$ is the flux density of the Crab Nebula, which depends on the antenna aperture and the observation epoch ({MJD}). 
Following \cite{2016ApJ...832..212M} (see also The flux density of the Crab Nebula subsection), 
we calculated $S_{\rm CN, U}=649 \pm 13 f({\rm MJD})$ Jy for Usuda and $S_{\rm CN, K}=717 \pm 14 f({\rm  MJD})$ Jy for Kashima,
where $f({\rm MJD})$ $\equiv \exp(-4.58\times 10^{-6} (\mbox{\rm  MJD}-57974))$
is a decaying factor of the total nebula emission. 
We included the decaying factor for completeness. However, this factor gives negligible $-0.3\%$ changes of $S_{\rm CN, U}$ and $S_{\rm CN, K}$
throughout observing sessions 1-15.

Our GRP selection criterion of SNR$_{\rm thr}=$5 corresponds to 
109$\pm 2$ 
Jy for Usuda and 
199$\pm 8$ 
Jy for Kashima on MJD$=$57974.
We calculate the fluence of a GRP from $I(t_k)$ according to
\begin{equation}
F= \sum_{t_k} I(t_k) \Delta t,
\label{eqn:CalFluence}
\end{equation}
where $t_k$ runs over  time intervals
during which we judge that there is  an enhancement of SNR($t_k$) relating to this GRP.
For example, the MP GRP shown in Fig. \ref{figure:A1} has 
a peak intensity of 3.7$\times 10^{4}$ Jy and fluence of 4.1$\times 10^{5}$ Jy \textmu s.

Figs. \ref{figure:A7} and \ref{figure:A8} show the differential flux and fluence histograms of the MP GRPs and IP GRPs 
for the combined data (Usuda$+$Kashima).
The middle part of the  intensity histograms, 
specifically in  intensity ranges $10^{2.9}$--$10^{4.2}$ Jy and $10^{2.7}$--$10^{3.5}$ Jy for MP and IP GRPs, respectively, 
are found to have power-law dependence on the intensity 
with indices of $-2.90\pm 0.19$ and 
$-2.56\pm 0.41$ (Fig.~\ref{figure:A7}).
Similarly, the middle part of the fluence histograms show power-law dependence (Fig. \ref{figure:A8}):  
fluence ranges of 
$10^{3.9} $--$10^{5.2}~{\rm Jy}~$\textmu s 
and $10^{3.7}$--$10^{4.5}~{\rm Jy}~$\textmu s, respectively, 
with indices of $-3.02\pm 0.19$ and $ -2.61 \pm 0.28$.
The spectral index for the MP GRP fluences is consistent with previous observations \cite{2011ARep...55..724Z} in the 2 GHz band.
These authors
reported the power-law index of $-2.0$ for
a cumulative fluence histogram of MP GRPs in  a fluence range of $10^{3.8}$--$ 10^{4.4}~{\rm Jy}~$\textmu s, 
which corresponds to the index $-$3.0 for the corresponding differential fluence histogram.
They also observed  an index change from $-$2.0 to $-$1.1 at a cumulated flux range below 
$10^{3.8}~{\rm Jy}~$\textmu s,
where our result is inconclusive  because it is close to  our detection limit.

\subsubsection*{The flux density of the Crab Nebula}\label{suppl:A-9}
With the angular distance from the center of Crab Nebula $\theta$, 
the spatial intensity distribution of the Crab Nebula is approximated as \cite{2016ApJ...832..212M}
\begin{equation}
B_{\rm CN}(\theta) = B_0 \left ( 1- \frac{\theta}{\theta_{\rm CN}} \right )
~~~~~~~~~~~~~~~~(0 \le \theta \le \theta_{\rm CN}),
\end{equation}
where size $\theta_{\rm CN}$ is in units of 3 arcmin.
The coefficient $B_0$ is normalized such that
\begin{equation}
2\pi \int^{\theta_{\rm CN}}_0 B_{\rm CN} (\theta)~\theta d\theta = S_{\rm CN, tot},
\end{equation}
where the total flux density of the Crab Nebula at  frequency $\nu_c$ is a decaying function at epoch 
$Y_{\rm obs}$ (year or MJD) \cite{2010ApJ...711..417M},
\begin{equation}
\begin{array}{ccll}
S_{\rm CN, tot}
&=& (973\pm19\mbox{ Jy})\displaystyle{\left ( \frac{\nu_c}{\rm 1~GHz} \right )^{-0.296\pm 0.006} }
             & \times \exp \left ( -1.67\times 10^{-3} ( Y_{\rm obs}-2003 ) \right )\\ \\
~~~~~~~~~~~~~~~ 
&=& (747\pm15\mbox{ Jy})\displaystyle{\left ( \frac{\nu_c}{2.25~{\rm GHz}} \right )^{-0.296\pm 0.006} }
             & \times \exp \left(-4.58\times 10^{-6} ({\rm MJD} - 57974) \right ). 
\end{array}
\label{eqn:DecayingScn}
\end{equation}
An antenna with a diameter $D$ at an observing wavelength $\lambda$ 
detects the Crab Nebula with a flux density
\begin{equation}
S_{{\rm CN}~D, \lambda} = 
\frac{\displaystyle{\int^{\theta_{\rm CN}}_0 B_{\rm CN} P(\theta; D, \lambda) (\theta)~\theta d\theta}} 
     {\displaystyle{\int^{\theta_{\rm CN}}_0 B_{\rm CN} (\theta)~\theta d\theta}}
     S_{\rm CN, tot},
\label{eqn:angularspread}
\end{equation}
where $P(\theta; D, \lambda)$ is the antenna pattern function
\begin{equation}
P(\theta; D, \lambda) =\exp \left [ -4 \ln 2 \left ( \frac{\theta}{\theta_b} \right )^2  \right]
\end{equation}
with beam size  $\theta_b=\lambda / D$ \cite{2013tra..book.....W}. 
At the center of our observing frequency $\nu_c=2.25$~GHz ($\lambda=13.3$~cm), 
we calculate $S_{\rm CN, U}$ for Usuda ($D=64$~m) and $S_{\rm CN, U}$ for Kashima ($D=34$~m)
from equation (\ref{eqn:angularspread}) and find
\begin{equation}
\begin{array}{ccc}
S_{\rm CN, U} &=& (649 \pm 13\mbox{ Jy})~f({\rm MJD}) \\ \\
S_{\rm CN, K} &=& (717 \pm 14\mbox{ Jy})~f({\rm MJD}),
\end{array}
\end{equation}
where $f({\rm MJD})=\exp \left(-4.58\times 10^{-6} ({\rm MJD} - 57974) \right )$
is the decaying factor from the second line of equation (\ref{eqn:DecayingScn}).

%% file: suppl_03_xray_v2.tex
\subsection*{X-ray data and analysis}
\label{suppl:xray}

\subsubsection*{Data reduction and verification} 
The X-ray timing instrument of {\it NICER} \cite{2016SPIE.9905E..1HG,Miller_2019} consists of co-aligned X-ray concentrators coupled with silicon drift detectors at its focal plane. 
The {\it NICER} detectors have a time resolution of $<100$ ns, corresponding to $<3\times10^{-6}$ cycles of the spin period of the Crab Pulsar.  It has 56 focal-plane modules,  four of  which were inactive during the observations. Figure~\ref{fig:effective_area}  
shows the {\it NICER}'s effective collecting area as a function of energy around 1.5 keV.

We used 15 {\it NICER} observations carried out in 2017--2019.  Table \ref{tab:nicer_log} summarizes the observation log.  
We performed basic data analysis with NICERDAS version 6 in HEAsoft 6.26 \cite{1999ascl.soft12002B} and {\it NICER} calibration database version 20190516. We created level-2 cleaned event files by applying \texttt{nicerl2} command to the unfiltered data, setting the \texttt{underonly\_range="0-500"} (default value: 200) and \texttt{overonly\_range="0-10"} (default value: 1) to maximize the available exposure time. To correct the photon arrival times to the barycenter of the Solar System, we used the \texttt{barycorr} command, adopting the coordinates of Crab  of R.A.=83.633218 and Decl.=22.014464 (J2000 equinox) and the Jet Propulsion Laboratory Solar System development ephemeris DE430 \cite{2014IPNPR.196C...1F}. We refer to the spin ephemerides listed in Table~\ref{tbl:A2} at the radio data and analysis section. The proper motion of the Crab Pulsar is negligible \cite{Kaplan_2008}. We created Good Time Intervals (GTIs) during which Crab was simultaneously observed with {\it NICER} and radio telescopes. All the X-ray photons that  were not covered by the merged GTI  were excluded in this analysis. 

Figure \ref{fig:crabspec}a shows the X-ray spectrum of the Crab Pulsar plus the surrounding nebula obtained with {\it NICER}. Compared  with the source spectrum, the background is negligible except for the lowest energy band. We found that the signal-to-noise ratio  was higher than 100 in 0.3--10 keV. Therefore, we selected X-ray photons in the energy range of 0.3--10 keV to maximize the signal to noise ratio. In the 0.3--10~keV band, the background contamination is $<$1\% of the source intensity. 
Figure~\ref{fig:crab_pulse_dection_significance}  shows the detection significance with  {\it NICER}  of  a pulsation from the Crab Pulsar as a function of the exposure time. The Crab pulsation  is detectable at an exposure of $\sim$1~sec  ($\sim$30 pulse rotations) or longer. Movie S1 shows how the accumulated Crab pulse profile  evolves as the {\it NICER} exposure increases. 

Figure~\ref{fig:profile_polyfit} shows the X-ray pulse profile around the MP of the Crab Pulsar obtained by stacking all the {\it NICER} data. We folded the light curve with 8192 phase bins, and decompose the profile with the Fourier transform. The pulse peak is measured at $\phi=$0.99125$\pm$0.00004, which is estimated by the Fourier series with 100 harmonics \cite{2003A&A...411L..31K}. We also fitted 
the pulse profile near the X-ray peak with a 4-th order polynomial function, which gave consistent results to 
the Fourier-filtered pulse profile. This corresponds  to  a radio delay of  $\sim$304~\textmu s compared to the X-rays; this delay is consistent with the previously reported ranges [figure 2 of \cite{2004ApJ...605L.129R}, figure~4 of \cite{2003A&A...411L..31K}, and  \cite{2010ApJ...708..403M}]. The stability of the MP peak is shown in Figure~\ref{fig:xray_center}. The uncertainties of the peak determination and phase jump of the {\it NICER} monitoring are smaller than those obtained in previous studies. 

\subsubsection*{Search for X-ray enhancement coinciding with GRPs}
\label{suppl:search_for_enhancement}
We assigned a pulse phase to each X-ray photon  on the basis of the radio ephemeris in Table~\ref{tbl:A2}, and classified them into three categories: MP-GRP associated, IP-GRP associated, and GRP unassociated events. Given that the MP-GRP arrival phases were distributed between pulse phases 0.992 and 1.008, we define  an MP-GRP associated cycle as  one where the main peak  encompasses MP-GRPs. We calculated the pulsed emission, using the count rate between phases 0.985 and  0.997. This phase interval, 0.012 spin cycles centered at the X-ray peak, is selected for comparison with previous optical measurements \cite{2003Sci...301..493S,2013ApJ...779L..12S}. The off-pulsed emission between phases 0.7 and 0.8, which included emission from the pulsar wind nebula and background, was subtracted from the light curve. The enhancement  was calculated from the difference between the averaged pulsed emission from the MP-GRP associated  and  GRP-unassociated cycles.

The detection significance of the enhancement  was estimated with Monte Carlo simulations (see Figure~\ref{fig:simulated_histogram}a).
In each simulation run, we randomly choose 24851 cycles, which contains the same amount as that of the MP-GRP-associated one, from the non-GRP-associated sample. Then, we treated these selected cycles as fake MP-GRP-associated cycle and calculated the enhancement as $p_i$ using the same way as in the previous paragraph. We repeated this procedure $10^4$ times. The mean value $p_{\textrm{sim}}$ of all $p_i$ is expected to be close to zero because the sample is chosen from non-GRP-associated cycles. The standard deviation of the enhancement of the simulated data sets $\sigma_{\textrm{sim}}$ represent 1$\sigma$ significance level if the observed enhancement is caused by the statistical fluctuation. The detection significance is defined as 
\begin{equation}
    \textrm{significance}=\frac{p_{\textrm{MP-GRP}}-p_{\textrm{sim}}}{\sigma_{\textrm{sim}}},
\end{equation}
where $p_{\textrm{MP-GRP}}$ is the observed enhancement from MP-GRP associated cycles. The resulting X-ray enhancement is described in the main text. 

In the main text, we fixed the phase width at 0.012 and center at 0.991 to compare our X-ray results with the previous reports at the optical wavelength \cite{2013ApJ...779L..12S}. Figure~\ref{fig:width_vs_enhancement} shows the pulse phase in the MP  during which the X-ray enhancement appeared as a function of the pulse phase center. The constant enhancement stays at $\sim 3$\% level in the $\phi \sim 0.97-1.00$ range although the significance is lower than 5$\sigma$. This suggests that  the X-ray enhancement occurs at a wider phase range than the range ($\delta \phi = 0.012$) found in the optical.

We further divided MP-GRPs into two groups with $\phi_{\rm{GRP}}<0$ and $\phi_{\rm{GRP}}\geq0$, and tested whether the associated X-ray enhancements are GRP-phase-dependent. Following the same procedure described above, the X-ray pulse profiles associated with these two groups of MP-GRPs are shown in Figure \ref{fig:profile_diff_grp_phase}. 
To test whether the enhancement is GRP-phase-dependent, we enlarged the phase range of calculation to 0.975--1.000. We found an enhancement of $2.8\pm0.8$ \% with a significance of 3.5$\sigma$ for $\phi_{\rm{GRP}}<0$ and an enhancement of $3.3\pm0.7$ \% with a significance of 4.5$\sigma$ for $\phi_{\rm{GRP}}>0$. The difference in the enhancements of these two groups is not statistically significant. We also divided MP-GRPs into bright and faint groups according to their radio fluxes, but did not find a difference between the X-ray enhancement associated with bright and faint MP-GRPs. 

Finally, we searched for enhancement of IP-GRP associated cycles. Following above procedures, the pulse profiles with GRP unassociated cycles and IP-GRP associated cycles near the IP are shown in Figure 1
Compared to Figure 1, the uncertainties of IP-GRP associated pulse profile are large and we could not observe statistically significant enhancement and  estimated a 3$\sigma$ upper limit of the enhancement of $\sim10$\%.

\subsubsection*{Lag analysis}
\label{suppl:lag_analysis}
We also performed a lag analysis by shifting the pulse number of X-ray events to search for potential correlations around the MP. The results of this analysis are shown in Figure~\ref{fig:simulated_histogram}b for pulse lags of $\pm 10$ cycles. We found no statistically significant enhancement ($>5$ sigma) around the lag over $\pm$10 cycles and thus confirmed our detection.

\subsubsection*{Spectral analysis for the MP-GRP-associated X-rays}
\label{suppl:spectral_study}

For the spectral analysis, we extracted the MP-GRP-associated and non-GRP-associated spectra as described in the main text. Events  were selected within the phase interval $0.985<\phi<0.997$ (see Search for X-ray enhancement coinciding with GRPs subsection~, 
i.e., with the phase width of $\delta\phi=0.012$). For the background spectra, we extracted events from the phase interval $0.700<\phi<0.800$, i.e., with  $\delta \phi = 0.100$ as described in the main text.  The exposure times for the MP-GRP-associated and non-GRP-associated spectra were  calculated by taking the number of unique rotations  encompassing the extracted MP-GRP-associated and non-GRP-associated events, respectively, then multiplying by the corresponding spin frequencies and by $\delta \phi$.  This spectral extraction procedure was performed for individual ObsIDs. The individual spectra were then combined with the HEASoft \texttt{mathpha} tool \cite{1999ascl.soft12002B}, which merges source and background spectra. The source spectra are binned so that each spectral bin has at least 50 photons. 
We used xspec version 12.11.1 \cite{arnaud96}  for the spectral analyses of the combined MP-GRP-associated  and  non-GRP-associated spectra. We used the response matrix file (``nixtiref20170601v002.rmf") and auxiliary response file (``nixtiaveonaxis20170601v004.arf") released on July 22, 2020. We included 0.5\% systematic uncertainty into the Non-MP-associated spectrum to allow for the uncertain accuracy of the detector response, which is known to be more pronounced for brighter objects.  

The spectral model employed (\texttt{tbfeo$\times$logpar}) was a log-parabola-shaped power-law (\texttt{logpar}) with an energy-dependent index \cite{ge12} convolved with the absorption with letting the oxygen and iron abundances free (\texttt{tbfeo}) \cite{wilms00}. Only the absorption parameters, the absorption column density $N_{\rm H}$ and abundances for oxygen and iron, were tied between the MP-GRP-associated and non-GRP-associated spectra --- the other model parameters  for the two spectra were free parameters.  We used the 0.2--10.0 keV range. The  \texttt{logpar} model is defined by the X-ray flux $A(E)$ (photons~keV$^{-1}$~cm$^{-2}$~s$^{-1}$) as a function of an X-ray energy $E$ as 
\begin{eqnarray}
A(E)= K \left(E/E_{\rm pivot}\right)^{\left(-\alpha-\beta \log \left(E/E_{\rm pivot}\right) \right)}, 
\end{eqnarray}
where $\alpha$, $\beta$, and $K$ (photons~keV$^{-1}$~cm$^{-2}$~s$^{-1}$) are the  parameters to be determined, and $E_{\rm pivot}=1$~keV. 
 Table~\ref{tab:spec_results} lists the best-fitting parameters for this model with 1$\sigma$-confidence uncertainties. The spectral parameters $\alpha$ and $\beta$ of the two spectra  were consistent within $1\sigma$ confidence. 
 However, the 0.2-12 keV flux of the MP-GRP-associated spectrum, $(2.25\pm 0.04)\times 10^{-8}$~erg~s$^{-1}$~cm$^{-2}$, was $3.6$\% higher than that of the non-GRP-associated spectrum, $(2.171\pm 0.004)\times 10^{-8}$erg~s$^{-1}$~cm$^{-2}$. 
 If we allow only the normalization parameters free with fixing the other parameters, the 0.2--12~keV flux is $(2.25\pm 0.01)\times 10^{-8}$ erg~s$^{-1}$~cm$^{-2}$ and $(2.171\pm 0.003)\times 10^{-8}$ erg~s$^{-1}$~cm$^{-2}$ for the MP-GRP-associated and non-GRP-associated spectra, respectively. 
 This corresponds with the enhancement at $3.6\pm0.2$\%, statistically consistent with the enhancement detected from the X-ray pulse profile in the main text.

We also investigated the cumulative distributions of the MP-GRP-associated and non-GRP-associated counts as a function of the instrument channel (i.e., free of  uncertainty of the instrumental response).  A Kolmogorov--Smirnov test on the exposure-scaled cumulative distributions demonstrates that both spectra follow the same count distribution, i.e., that they have the same spectral shape. This corroborates the finding of the spectral analyses above.  

We extracted the net enhanced X-ray spectrum using the Non-GRP-associated data as background subtracted from the MP-GRP-associated spectrum. Due to the low statistics of this differential spectrum, we can not determine the absorption parameters. Using the same \texttt{tbfeo} model with the parameters fixed at the values shown in Table~\ref{tab:spec_results}, the incident continuum is approximated ($\chi^2$/d.o.f.=8.89/7) by a single power-law model with photon index $\Gamma=1.66_{-0.49}^{+0.68}$ and an absorbed 0.2-12 keV flux of $8.37^{+2.98}_{-2.96}\times 10^{-10}$~erg~s$^{-1}$~cm$^{-2}$, which is consistent with the differential flux estimated from  Table~\ref{tab:spec_results}. This is shown in Figure~\ref{fig:netfit_spec}. This differential spectrum can be also approximated by the  \texttt{tbfeo$\times$logpar} model with the same (fixed) $\alpha$ and $\beta$ values ($\chi^2$/dof=9.74/8) only by changing its normalization. Thus, we can not statistically distinguish the MP-GRP-associated spectral shape from that of the regular pulses.

Figure~\ref{fig:SED_suppl} shows the broad-band spectral energy distribution of the flux enhancement of the Crab Pulsar  during  an MP-GRP.  The X-ray enhancement was 3.8\% of the persistent flux  and showed  no spectral change from the  persistent emission.

%% file: suppl_04_theorical_v2.tex
\subsection*{Theoretical models for the X-ray enhancement during GRPs} 
\label{suppl:theory}

\subsubsection*{Simple emission model from bunching particles}
\label{suppl:particle_bunches}
One simple interpretation of the radio and X-ray enhancements is a temporary increase
of particle numbers in the emitting region. Only a small fraction $V_{\rm radio}/V_{\rm X}$
of the X-ray emitting volume $V_{\rm X}$ would be linked to the radio emitting region,
which is common for both regular pulses and GRPs.
This is supported by the narrow phase where the X-ray enhancement is detected (Figure \ref{fig:pulse_profile}),
although the estimate of the actual volume of the emission would not be simple due to the caustic nature of the emitting region (i.e., emission originating at different regions in the magnetosphere being piled-up in phase, \cite{1983MNRAS.202..495M}). 
If coherent radio emission comes from particle bunches, the flux for a fixed number of bunches is proportional to the square of the particle density (e.g., \cite{1975A&A....44..285S}). 
Then a density increase of $\sim$16 times in the radio emitting volume $V_{\rm radio}$ leads to the typical flux ratio of GRP to regular radio pulses of $16^2 \simeq 2.5 \times 10^2$. 
On the other hand, the enhancement of the X-ray flux with the emitting volume $V_{\rm X}$ is $\sim3.8\%$. This suggests the volume fraction of the radio emitting region $V_{\rm radio}/V_{\rm X} \sim 0.038/16 \simeq 2 \times 10^{-3}$.

\subsubsection*{Magnetic reconnection model}
\label{suppl:reconnection}

We estimate the X-ray luminosity of GRP-emitting plasmoids on the basis of a reconnection model \cite{2019ApJ...876L...6P}. 
We show a schematic picture of the model in Figure \ref{fig:interpretation}. 
In this model, a GRP is produced by collisions of multiple plasmoids in the current sheet beyond the light cylinder. 
Because the current sheet is unstable to plasmoid instability, the sheet is fragmented into a dynamical chain of plasmoids due to the magnetic reconnection [e.g., \cite{2017A&A...607A.134C}].
A collision of plasmoids ejects fast magnetosonic waves, which escape from the plasma as electromagnetic radio waves. 
Considering cross-layer pressure balance and energy balance between heating by the magnetic energy dissipation and synchrotron cooling, the plasma density\ $n$, thermal Lorentz factor\ $\gamma_{\rm th}$, and current sheet thickness $\delta_{\rm cs}$ in the comoving frame are estimated to be $n\sim7\times10^{11}(B_{\rm LC}/10^6{\rm G})^{5/2}$ cm$^{-3}$, $\gamma_{\rm th}\sim8\times10^4(B_{\rm LC}/10^6{\rm G})^{-1/2}$, and $\delta_{\rm cs}\sim10^2(B_{\rm LC}/10^6{\rm G})^{-3/2}$ cm, respectively, for  the magnetic field $B_{\rm LC}\sim10^6$ G at the light cylinder of the Crab Pulsar \cite{2014ApJ...780....3U}. 
Numerical simulations of the reconnection process \cite{2019ApJ...876L...6P} show that the typical size of a plasmoid is $\sim10-100$ times larger than the thickness of the current sheet. 
As a result, the amount of magnetic energy released in an individual plasmoid collision is
\begin{eqnarray}
E_{\rm B}&\sim&(B_{\rm LC}^2/8\pi)l^3 \nonumber \\
&\sim&4\times10^{22}\left(\frac{B_{\rm LC}}{10^6~{\rm G}}\right)^2\left(\frac{l}{10^4~{\rm cm}}\right)^3~{\rm erg}, 
\end{eqnarray}
where $l\sim100\delta_{\rm cs}$ is  the typical size of a plasmoid. 
Assuming that a fraction ($\sim0.01$) of the released magnetic energy of a plasmoid is converted to radio emission \cite{2019ApJ...876L...6P} and that the bulk Lorentz factor of  the plasma flow in the current sheet is $\Gamma\sim100$ (e.g.,  \cite{1996A&A...311..172L}), the observed flux density of an individual plasmoid merger event is expected to be 
\begin{eqnarray}
S_{\rm pl}&\sim&\frac{0.01E_{\rm B}}{\pi d^2(l/c)\nu}\Gamma^3 \nonumber \\
&\sim&5\times10^2\left(\frac{B_{\rm LC}}{10^6~{\rm G}}\right)^2\left(\frac{l}{10^4~{\rm cm}}\right)^2\left(\frac{\nu}{2~{\rm GHz}}\right)^{-1}\left(\frac{\Gamma}{10^2}\right)^3~{\rm Jy}, 
\end{eqnarray}
where $d\sim2$ kpc is the distance to the Crab Pulsar \cite{1973PASP...85..579T}. 
 Given that a large number of plasmoids would merge near the light cylinder of the Crab Pulsar due to  tearing instability \cite{2019ApJ...876L...6P}, multiple nanoshots may form a single GRP
(e.g., \cite{2016JPlPh..82c6302E}). 
In our observations, the typical duration of a  GRP is $t_{\rm dur}\sim1.5\times10^{-5}$ s and the fluxes of most of the detected GRPs  are close to the observation threshold value of $S_{\rm th}\sim100$ Jy at 2 GHz (see the power-law distribution in Figure \ref{figure:A6}). 
Then, the number of the merging plasmoids observed in a  GRP is estimated to be 
\begin{eqnarray} 
N_{\rm pl}&\sim&\left(\frac{t_{\rm dur}}{(l/c)/\Gamma}\right)\left(\frac{S_{\rm th}}{S_{\rm pl}}\right) \nonumber \\
&\sim&10^3\left(\frac{B_{\rm LC}}{10^6{\rm G}}\right)^{-2}\left(\frac{l}{10^4~{\rm cm}}\right)^{-3}\left(\frac{\nu}{2~{\rm GHz}}\right)\left(\frac{\Gamma}{10^2}\right)^{-2}.
\end{eqnarray}

The GRP-emitting plasmoids could emit synchrotron radiation in X-ray. 
Heating in the current sheet may be highly variable. A large fraction of particles has a Lorentz factor lower than the average value of $\gamma_{\rm th}$. 
We assume that there is a fraction $\eta_{\rm X}$ of low-energy electrons, which emit synchrotron radiation with the characteristic energy $h\nu_{\rm X}\sim1$ keV, corresponding to the Lorentz factor $\gamma\sim20(B/10^6{\rm G})^{-1/2}$, in the GRP-emitting plasmoid. 
Then the synchrotron luminosity of a group of the multiple plasmoids which emit a single GRP is calculated to be 
\begin{eqnarray}
L&\sim&\frac{4\pi}{3}\eta_{\rm X}nl^3N_{\rm pl}\frac{2e^4B^2\gamma^2}{3m_{\rm e}^2c^3}\Gamma^4 \nonumber \\
&\sim&3\times10^{34}\left(\frac{\eta_{\rm X}}{0.3}\right)\left(\frac{B_{\rm LC}}{10^6{\rm G}}\right)^{3/2}\left(\frac{\nu}{2~{\rm GHz}}\right)\left(\frac{h\nu_{\rm X}}{1~{\rm keV}}\right)\left(\frac{\Gamma}{10^2}\right)~{\rm erg~s}^{-1}.
\end{eqnarray} 
This luminosity corresponds to $\sim3$\% of the average X-ray luminosity of the Crab Pulsar, which indicates that the observed radio and X-ray enhancements can be explained by the emission from plasmoids in the reconnecting current sheet beyond the light cylinder.

\subsubsection*{Radio absorption model}
\label{suppl:absorption}

An alternative model for the X-ray flux enhancement correlated with a GRP is the cyclotron resonant absorption of radio photons by the synchrotron-emitting particles \cite{LyubarskiPetrova1998}.  In this model,  relativistic particles maintain their pitch angles by absorbing radio photons that are in the cyclotron resonance in their rest frame. The power for the X-ray emission comes from the energy of the particles with  radio photons acting as  catalysts, providing the relativistic particles with  a certain pitch angle so that they radiate synchrotron emission. The particles start out with very small pitch angles.  When they begin absorbing the radio emission, their perpendicular momenta increase stochastically until they reach high Landau states, where their losses in the perpendicular momenta  and increases in the pitch angles from absorption balance out.
The resonant condition will occur in the outer magnetosphere near and beyond the light cylinder for the Crab Pulsar \cite{2008ApJ...680.1378H}.  The X-ray synchrotron flux should, therefore,  depend on the radio flux that can be absorbed.  Some of the radio  emission in the direction of the observer may be absorbed, and the magnitude of the absorption optical depth is discussed elsewhere \cite{Petrova 2002}.

We estimate the expected amount of the increase of the X-ray flux  from the  increase of the radio flux during a GRP.  Assuming that the momentum of  a particle perpendicular to the magnetic field has reached an equilibrium between absorption and radiation losses [\cite{Harding2005}, their equations (29) and (30)], we have the flux of synchrotron radiation
\begin{equation}
    F_{\rm syn} \propto \phi_0^{1/2},
\end{equation}
where $\phi_0$ is the radio flux density.  If the GRP flux is, on average, 100 times the regular radio flux, then the model would predict that the total synchrotron flux would increase by a factor of $\sim 10$.  However, the X-ray flux in the NICER band makes up only a fraction of the total Crab synchrotron flux, which extends from UV to gamma rays, and hence the expected flux increase in the NICER band would be much smaller.   The model predicts an enhancement of the synchrotron component in the other energy bands as well and most of the power in the Crab spectrum lie between 50 keV and 1 MeV. No significant constraints on the enhancement in the MeV band have been reported.